\documentclass[12pt]{article}
\usepackage{a4wide,amssymb,cite}
\pdfoutput=1

\usepackage{a4wide,amssymb,graphicx}
\usepackage{epsfig}
\usepackage[usenames,dvipsnames]{color}
\usepackage{slashed}
\usepackage{amssymb,cite,graphicx}
\usepackage{slashed}
\usepackage{amsmath,bm,bbm}
\usepackage{amsfonts}
\usepackage[titletoc,title]{appendix}
\usepackage[small]{caption}
\usepackage[margin=1in]{geometry}
\usepackage[multiple]{footmisc}
\usepackage{mathtools}
\usepackage{slashed}
\usepackage[nottoc]{tocbibind}
\usepackage{xcolor}

\newcommand{\be}{\begin{equation}}
\newcommand{\ee}{\end{equation}}
\newcommand{\bea}{\begin{eqnarray}}
\newcommand{\eea}{\end{eqnarray}}

\def\circa#1{\,\raise.3ex\hbox{$#1$\kern-.75em\lower1ex\hbox{$\sim$}}\,}

\begin{document}

\begin{titlepage}
%
%


%

\begin{centering}
\vspace{1cm}
{\Large {\bf Flux-mediated Dark Matter}} \\

\vspace{1.5cm}

{\bf Yoo-Jin Kang$^{\dagger}$, Hyun Min Lee$^{*}$, Adriana G. Menkara$^\star$ and Jiseon Song$^{\ddagger}$}
\vspace{.5cm}

{\it  Department of Physics, Chung-Ang University, Seoul 06974, Korea.}

\end{centering}
\vspace{2cm}

\begin{abstract}
\noindent
We propose a new mechanism to communicate between fermion dark matter and the Standard Model (SM) only through the four-form flux. The four-form couplings are responsible for the relaxation of the Higgs mass to the correct value and the initial displacement of the reheating pseudo-scalar field from the minimum. We show that the simultaneous presence of the pseudo-scalar coupling to fermion dark matter and the flux-induced Higgs mixing gives rise to unsuppressed annihilations of dark matter into the SM particles at present, whereas the direct detection bounds from XENON1T can be avoided.
We suggest exploring the interesting bulk  parameter space of the model for which dark matter annihilates dominantly into a pair of singlet-like scalars with similar mass as for dark matter.

\end{abstract}

\vspace{3cm}

\begin{flushleft} 
$^\dagger$Email: yoojinkang91@gmail.com  \\
$^*$Email: hminlee@cau.ac.kr \\
$^\star$Email: amenkara@cau.ac.kr \\
$^\ddagger$Email:  jiseon734@gmail.com
\end{flushleft}

\end{titlepage}

\section{Introduction}

Weakly Interacting Massive Particles (WIMPs) have been the main paradigm for particle dark matter for the last four decades or so. It is typical that the standard thermal freeze-out mechanism for WIMP dark matter depends on sizable interactions between dark matter and the SM particles, but doubts have been cast on the WIMP paradigm in view of the strong limits from direct detection experiments such as  XENON1T  \cite{xenon1t}, LUX \cite{lux}, PandaX-II \cite{panda}, etc.  On the other hand, there are potentially interesting excesses or signatures for indirect detection of WIMP dark matter in cosmic ray experiments, such as Fermi-LAT \cite{fermi-lat}, HESS \cite{hess}, AMS-02 \cite{ams}, etc.

Recently, the interesting possibility to relax the Higgs mass and the cosmological constant to right values through the four-form flux has been revisited \cite{hierarchy,Giudice,Kaloper,hmlee1,hmlee2,hmlee3}. A dimensionless four-form coupling to the Higgs field makes the Higgs mass variable until the observed small cosmological constant is achieved due to the last membrane nucleation \cite{nucl,polchinski}. But, in this scenario, the Universe would appear empty at the end of the last membrane nucleation without a reheating mechanism, because the previously produced particles would have been diluted due to prolonged dS phases.

There have been ways suggested to reheat the Universe in models with four-form flux, such as the non-perturbative particle production in the time-dependent background during the last membrane nucleation \cite{Giudice} and the decay of an extra singlet scalar field whose potential has the flux-dependent minimum \cite{hmlee1,hmlee2,hmlee3}. In the former case, the particle production rate depends on the speed of transition for the last membrane nucleation, thus there would be a need of a small membrane tension for the efficient particle production \cite{Giudice}. On the other hand, in the latter case, the singlet scalar field has a sufficiently large latent heat after the membrane nucleation, so the perturbative decay of the singlet scalar field gives rise to an efficient reheating of the Universe \cite{hmlee1,hmlee2,hmlee3}.

In this article, we consider a Dirac fermion dark matter in models with a four-form flux and a singlet pseudo-scalar field.  Both the Higgs field and the singlet pseudo-scalar field couple directly to the four-form flux such that the Higgs mass as well as the vacuum expectation value (VEV) of the pseudo-scalar field are relaxed to true values at the same time.  Since dark matter has a direct coupling to the pseudo-scalar field, we can communicate between dark matter and the Standard Model (SM) particles only through the four-form couplings.  We dub this scenario ``Flux-mediated dark matter''.  

The simultaneous presence of the CP-odd four-form coupling to the pseudo-scalar field and the CP-even four-form coupling to the Higgs field gives rise to the CP violation in the dark sector. As a result, the pseudo-scalar coupling to dark matter and a flux-induced Higgs mixing lead to  unsuppressed dark matter annihilations into the SM at present whereas suppressing the elastic scattering cross section between dark matter and nucleons for direct detection.  We discuss the possibility of obtaining the observable signals for indirect detection such as in Fermi-LAT and AMS-02 while satisfying the correct relic density, the strong constraints from XENON1T and the other bounds from Higgs and electroweak data and collider searches.

The paper is organized as follows. 
We first present the model setup with the four-form flux, the pseudo-scalar field as well as dark matter.
 Then, we review the relaxation of the Higgs mass from the flux-dependent minima of the scalar potential and the reheating from the perturbative decay of the pseudo-scalar field in our model. Next we provide new results for flux-mediated dark matter and discuss the relic density of dark matter, the current bounds from direct and indirect detection of dark matter and Higgs/collider data. Finally, we show the combined constraints on the parameter space of our model and conclusions are drawn. There are two appendices summarizing the scalar self-interactions in our model and including the full formulas for dark matter scattering.

\section{The model}

For the scanning of the Higgs mass and the cosmological constant, we introduce a three-index anti-symmetric tensor field $A_{\nu\rho\sigma}$, whose four-form field strength is given by  $F_{\mu\nu\rho\sigma}=4\, \partial_{[\mu} A_{\nu\rho\sigma]}$. Moreover, we add a pseudo-scalar field $\phi$ for reheating after the relaxation of the Higgs mass and consider a Dirac singlet fermion $\chi'$ for dark matter  \footnote{We introduced the primed notation  $\chi'$ for fermion dark matter, and we reserve the unprimed notation $\chi$ for the physical basis later. }.  

\begin{figure}[tbp]
\centering 
\includegraphics[width=.55\textwidth]{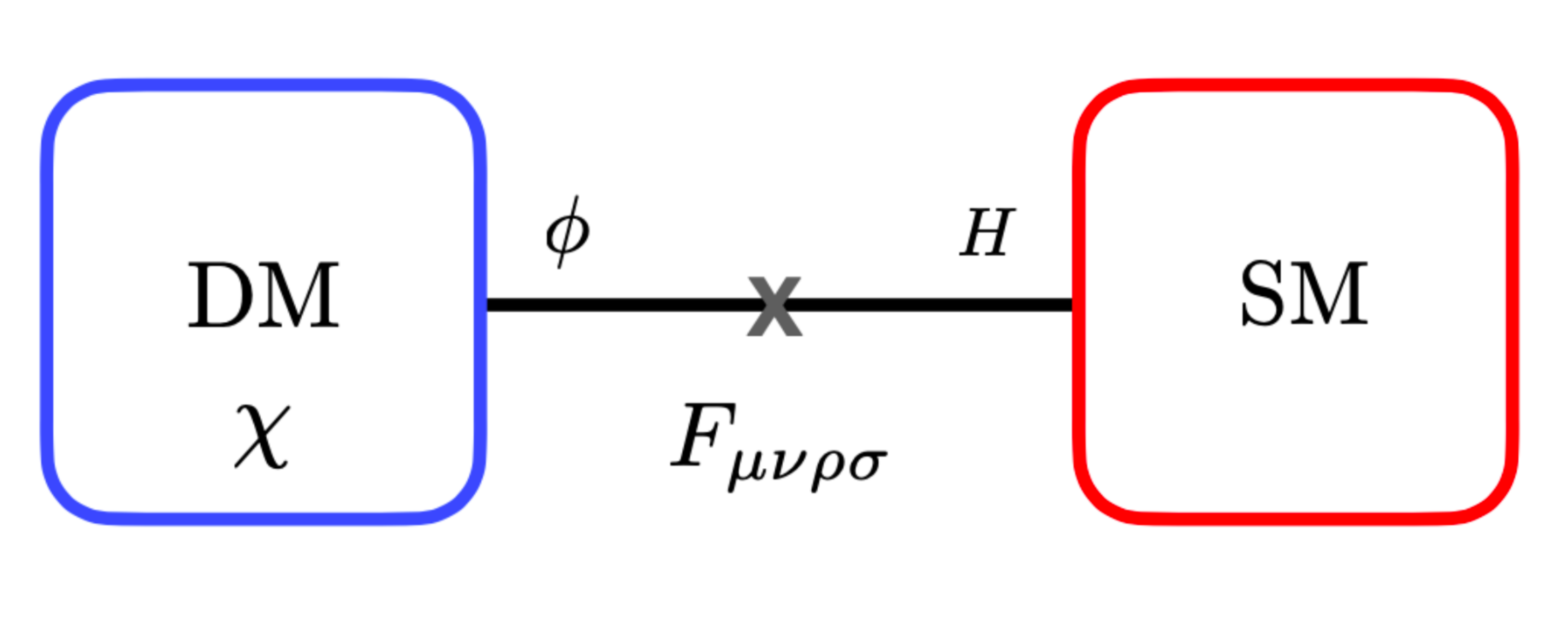} 
\caption{\label{fig:fmdm}  Schematic diagram for flux-mediated dark matter.  }
\end{figure}

We consider the Lagrangian with four-form field couplings included beyond the SM, which is composed of various terms as follows,
\bea
{\cal L} = {\cal L}_0 +{\cal L}_{\rm ext} \label{full}
\eea
where 
\bea
 {\cal L}_0 &=&  \sqrt{-g} \bigg[\frac{1}{2}R  -\Lambda -\frac{1}{48} F_{\mu\nu\rho\sigma} F^{\mu\nu\rho\sigma} \nonumber \\
 &&-  |D_\mu H|^2 -M^2 |H|^2 +\lambda_H |H|^4 +  \frac{c_H}{24} \,\epsilon^{\mu\nu\rho\sigma} F_{\mu\nu\rho\sigma} \, |H|^2 \nonumber \\
 &&  -\frac{1}{2} (\partial_\mu\phi)^2-\frac{1}{2} m^2_\phi(\phi-\alpha)^2 +\frac{\mu}{24} \,\epsilon^{\mu\nu\rho\sigma} F_{\mu\nu\rho\sigma} \, \phi \nonumber \\
&&+ i{\bar\chi}'\gamma^\mu \partial_\mu\chi'-m'_\chi {\bar\chi}'\chi'+ i\,\frac{m'_\chi}{f}\, \phi\, {\bar\chi}' \gamma^5 \chi' \bigg]. \label{L0}
\eea
and the extra Lagrangian ${\cal L}_{\rm ext}$ is composed of ${\cal L}_{\rm ext}={\cal L}_S+{\cal L}_L+ {\cal L}_{\rm memb} $ with
\bea
 {\cal L}_S &=&\frac{1}{6}\partial_\mu \bigg[\Big( \sqrt{-g}\,  F^{\mu\nu\rho\sigma} -c_H \epsilon^{\mu\nu\rho\sigma}  |H|^2 -\mu\,\epsilon^{\mu\nu\rho\sigma}\,  \phi   \Big)A_{\nu\rho\sigma} \bigg],  \\
 {\cal L}_L &=& \frac{q}{24}\, \epsilon^{\mu\nu\rho\sigma} \Big( F_{\mu\nu\rho\sigma}- 4\, \partial_{[\mu} A_{\nu\rho\sigma]} \Big),  \label{LL} \\
 {\cal L}_{\rm memb}&=& \frac{e}{6} \int d^3\xi\,  \delta^4(x-x(\xi))\, A_{\nu\rho\sigma} \frac{\partial x^\nu}{\partial \xi^a} \frac{\partial x^\rho}{\partial \xi^b} \frac{\partial x^\sigma}{\partial \xi^c} \,\epsilon^{abc} \nonumber \\
 &&-T\int d^3\xi\, \sqrt{-g^{(3)}}\,  \delta^4(x-x(\xi)).
\eea
 After a global $U(1)$ symmetry is broken spontaneously, $\phi$ could arise  as a pseudo-Goldstone boson and the mass term ($m'_\chi$) and the pseudo-scalar coupling ($m'_\chi/f$) for the fermion dark matter could be also generated.
We introduced $c_H$ and $\mu$ as dimensionless  and dimensionful couplings for the four-form flux to the SM Higgs \cite{hierarchy,Giudice,Kaloper,hmlee1,hmlee2} and the pseudo-scalar field \cite{inflation,hmlee3}, respectively.  We can take $c_H, \mu$  to be positive without loss of generality.   In our model, dark matter communicates with the SM through the four-form couplings. Thus, we dub  our scenario ``Four-form portals'' or ``Flux-mediated dark matter'' . We show the schematic diagram for flux-mediated dark matter in Fig.~\ref{fig:fmdm}.

The simultaneous presence of those four-form couplings to the Higgs and pseudo-scalar fields leads to the breakdown of the CP symmetry. In this case, we can avoid the direct detection bounds due to the velocity-suppression of DM-nucleon scattering but expect the indirect visible signals for dark matter at present, due to the unsuppressed pseudo-scalar coupling to fermion dark matter \footnote{See Ref.~\cite{axionmed,indirect} for fermion dark matter model with a pseudo-scalar mediator where the CP symmetry is unbroken. }. 

It is remarkable that the shift symmetry for the pseudo-scalar field is respected by the four-form coupling $\mu$ but it is softly broken by the mass term $m^2_\phi$ in the third line in eq.~(\ref{L0}). The soft-breaking mass $m_\phi$ for $\phi$ can be ascribed to a periodic potential, $\Lambda^{\prime 4}(1-\cos((\phi-\alpha)/F))$, with $\alpha/F$ being the arbitrary phase, which might be generated by a non-perturbative effect in the hidden sector.  In this case, we can identify the soft mass term by $m^2_\phi=\Lambda^{\prime 4}/F^2$ where $F$ could be different from $f$ appearing in the  axion-like coupling  of the Dirac fermion dark matter  $\chi$ to the pseudo-scalar field in the last line in eq.~(\ref{L0}).

We also comment that ${\cal L}_S$ is the surface term necessary for the well-defined variation of the action with the anti-symmetric tensor field, and $q$ in ${\cal L}_L$ (in eq.~(\ref{LL})) is the Lagrange multiplier, and $ {\cal L}_{\rm memb}$ contains the membrane action coupled to  $A_{\nu\rho\sigma}$ with membrane charge $e$ and the brane tension.
Here, $\xi^a$ are the membrane coordinates, $x(\xi)$ are the embedding coordinates in spacetime, $\epsilon^{abc}$ is the volume form for the membrane and $g^{(3)}$ is the determinant of the induced metric on the membrane.

Using  the equation of motion for $F_{\mu\nu\rho\sigma}$ \cite{hmlee1,hmlee2,hmlee3} as follows,
\bea
F^{\mu\nu\rho\sigma}=\frac{1}{\sqrt{-g}}\, \epsilon^{\mu\nu\rho\sigma} \Big(\mu\phi+ c_H |H|^2+q\Big),
\eea
and integrating out $F_{\mu\nu\rho\sigma}$, we recast the full Lagrangian (\ref{full}) into
\bea
{\cal L} &=&\sqrt{-g} \bigg[\frac{1}{2}R-\Lambda-  |D_\mu H|^2 +M^2 |H|^2 -\lambda_H |H|^4  \nonumber \\
&&\quad-\frac{1}{2} (\partial_\mu\phi)^2-\frac{1}{2} m^2_\phi(\phi-\alpha)^2-\frac{1}{2} (\mu \phi + c_H |H|^2+q)^2 \nonumber \\
&&\quad+ i{\bar\chi}'\gamma^\mu \partial_\mu\chi'-m'_\chi {\bar\chi}'\chi'+ i\,\frac{m'_\chi}{f}\, \phi\, {\bar\chi}' \gamma^5 \chi'\bigg] +{\cal L}_{\rm nucl}\label{Lagfull}
\eea 
with
\bea
{\cal L}_{\rm nucl}= \frac{1}{6}\epsilon^{\mu\nu\rho\sigma} \partial_\mu q A_{\nu\rho\sigma} +\frac{e}{6} \int d^3\xi \, \delta^4(x-x(\xi))\, A_{\nu\rho\sigma} \frac{\partial x^\nu}{\partial \xi^a} \frac{\partial x^\rho}{\partial \xi^b} \frac{\partial x^\sigma}{\partial \xi^c} \epsilon^{abc}.  \label{extra}
\eea
Then, the effective Higgs mass parameter, the effective cosmological constant and the effective Higgs quartic coupling are given by
\bea
M^2_{\rm eff}(q) &=& M^2 - c_H\, (q+\mu\langle\phi\rangle),  \label{effHmass}\\
  \Lambda_{\rm eff} (q) &=& \Lambda + \frac{1}{2}\, q^2 + V(\langle\phi\rangle)+V(\langle H\rangle), \\
  \lambda_{H,{\rm eff}}&=&\lambda_H+\frac{1}{2}c^2_H
  \eea
 where the Higgs mass induced by the VEV of the pseudo-scalar field and the vacuum energies coming from the Higgs and pseudo-scalar potentials are also included. 
 
 Moreover, the coupling between the pseudo-scalar and Higgs field is given by a direct product of  four-form couplings for them, $\mu$ and $c_H$, as can be seen from the expansion in eq.~(\ref{Lagfull}), mediating between dark matter and the SM. 
On the other hand, for scalar singlet dark matter $S$, we can introduce the four-form coupling to dark matter respecting a $Z_2$ symmetry by the interaction Lagrangian \cite{hmlee2}, $\frac{c_S}{24} \,\epsilon^{\mu\nu\rho\sigma} F_{\mu\nu\rho\sigma} \,S^2$. This results in the Higgs-portal coupling, $c_H c_S S^2 |H|^2$, similarly to the case with fermion dark matter. But, in this case, there is no reason to forbid the tree-level Higgs-portal coupling, $\lambda_{HS} S^2 |H|^2$, in the first place.
 This is in contrast to the case with fermion dark matter where the tree-level Higgs-portal coupling to the pseudo-scalar, $\mu_{\phi H}\phi |H|^2$, breaks the shift symmetry explicitly, thus it is forbidden.

On the other hand, the equation of motion for $A_{\nu\rho\sigma}$ in eq.~(\ref{extra}) makes the four-form flux $q$  dynamical, according to
\bea
\epsilon^{\mu\nu\rho\sigma} \partial_\mu q= -e\int d^3\xi \, \delta^4(x-x(\xi))\, \frac{\partial x^\nu}{\partial \xi^a} \frac{\partial x^\rho}{\partial \xi^b} \frac{\partial x^\sigma}{\partial \xi^c} \epsilon^{abc}.
\eea
The flux parameter $q$ is quantized in units of $e$ as $q=e\,n$ with $n$ being integer. 
As a result, whenever we nucleate a membrane, we can decrease the flux parameter by one unit  such that both the Higgs mass and the cosmological constant can be relaxed into observed values in the end. 

Before going into the details in the next section, we comment briefly on the relaxation of Higgs mass and cosmological constant. 
For $q>q_c$ with $q_c\equiv M^2/c_H-\mu\langle\phi\rangle$, the Higgs mass parameter in eq.~(\ref{effHmass}) becomes $M^2_{\rm eff}<0$, so electroweak symmetry is unbroken, whereas for $q<q_c$, we are in the broken phase for electroweak symmetry.  For $c_H={\cal O}(1)$ and the membrane charge $e$ of electroweak scale,  we obtain the observed Higgs mass parameter as $M^2_{\rm eff}\sim c_H\, e$, once the flux change stops at $q=q_c-e$ due to the suppression of a further tunneling with more membrane nucleation \cite{Giudice,Kaloper,hmlee1,hmlee2}.
For $\Lambda<0$, we can cancel a large cosmological constant by the contribution from the same flux parameter until $\Lambda_{\rm eff}$ takes the observed value at $q=q_c-e$, but we need to rely on an anthropic argument for that with $e$ being of order weak scale \cite{anthropic,Giudice}.  The detailed discussion on the vacuum structure and electroweak symmetry breaking will be discussed in the next section.

\section{Relaxation of Higgs mass and reheating}

We review the relaxation of the Higgs mass and the cosmological constant in the case with a singlet pseudo-scalar and discuss the reheating with four-form couplings.

\subsection{Flux-dependent minimum and Higgs mass}

For a general flux parameter $q$, we expand the SM Higgs and the pseudo-scalar  around the vacuum \cite{hmlee2} as $\langle H\rangle=(0,v_H(q)+h)^T/\sqrt{2}$ and $\langle\phi\rangle=v_\phi+\varphi$, with
\bea
v_H(q)&=& \sqrt{\frac{M^2- c_H(q+\mu v_\phi)}{\lambda_H+\frac{1}{2} c^2_H}},  \label{vev1} \\
v_\phi(q) &=&\frac{m^2_\phi}{\mu^2+m^2_\phi} \bigg[\alpha-\frac{\mu}{m^2_\phi} \cdot\Big( \frac{1}{2} c_H v^2_H + q\Big)\bigg]. \label{vev2}
\eea
The minimum of the potential is stable as far as $m^2_\varphi m^2_h> c^2_H \mu^2 v^2_H(q)$, where $m^2_\varphi=m^2_\phi+\mu^2$ and $m^2_h=2\lambda_{H,{\rm eff}} v^2_H(q)$. In the true electroweak minimum, we take the Higgs VEV  to $v_H(q_c-e)={\rm 246}\,{\rm GeV}$. 
Performing the following transformation to the mass eigenstates, $(h_1,h_2)^T$,
\bea
\left(\begin{array}{c}  h_1\\ h_2 \end{array} \right)= \left(\begin{array}{cc} \cos\theta(q) & -\sin\theta(q) \\ \sin\theta(q) &  \cos\theta(q) \end{array} \right) \left( \begin{array}{c}  \varphi \\ h \end{array} \right),
\eea 
we obtain the mass eigenvalues and the mixing angle $\theta(q)$ as
\bea
m^2_{h_{1,2}}=\frac{1}{2} (m^2_\varphi+m^2_h) \mp \frac{1}{2} \sqrt{(m^2_\varphi-m^2_h)^2+4c^2_H \mu^2 v^2_H(q)},  \label{masses}
\eea
and
\bea
\tan2\theta(q) = \frac{2c_H\mu v_H(q)}{m^2_\varphi-m^2_h}.  \label{mixing}
\eea
Then, we can trade off $c_H \mu$ for the Higgs mixing and the scalar masses. For a small mixing angle, $\theta\ll 1$, we can approximate $c_H \mu\approx \theta(q)\,(m^2_\varphi-m^2_h)\approx \theta(q) (m^2_{h_1}-m^2_{h_2})$, and $h_2$ is SM Higgs like and $h_1$ is pseudo-scalar like.
We find that even for a vanishing VEV of the pseudo-scalar, there is a nonzero mixing due to the four-form couplings.
Therefore, there is an one-to-one correspondence between the four-form coupling, $c_H\mu$, and the Higgs mixing angle, $\theta$, for given scalar masses. 

We note that in the absence of an explicit breaking of the shift symmetry, that is, $m^2_\phi=0$, there is no relaxation of a large Higgs mass, due to the fact that the minimization of the pseudo-scalar potential cancels the flux-induced Higgs mass completely. Thus, it is crucial to keep the explicit breaking mass term to be nonzero \cite{hmlee2}.

We also comment on the loop corrections and the naturalness of the pseudo-scalar field in our model. First, we find that the singlet-like scalar receives a logarithmically divergent mass correction at one-loop from the flux-induced coupling, ${\cal L}\supset - \frac{1}{2}c_H\mu \varphi h^2$, as follows, 
\bea
\delta m^2_\varphi= \frac{1}{64\pi^2} c^2_H \mu^2\, \ln \frac{\Lambda^2}{m^2_h}
\eea
where $\Lambda$ is the cutoff scale.
So, the mass correction is proportional to the pseudo-scalar mass, so it is technically natural to keep the singlet-like scalar light. 

Secondly, the four-form couplings lead to a quadratically divergent tadpole for the pseudo-scalar field by
$\Delta^3 \phi$ with $\Delta^3=\frac{c_H \mu\Lambda^2}{16\pi^2}$, 
which can be renormalized by the counter term $\alpha$ in eq.~(\ref{L0}).  The large tadpole term would result in a shift in the effective Higgs mass in eq.~(\ref{effHmass}), but it can be relaxed by the change of the four-form flux, because the effective tadpole term is given by $(\alpha m^2_\phi-\mu q-\Delta^3)\phi$. Otherwise, we could keep a small tadpole term technically natural by assuming a discrete symmetry with an extra Higgs-like scalar $H'$.  For instance, if the extra Higgs-like scalar has a four-form coupling of the opposite sign, ${\cal L}\supset-\frac{c_H}{24}\,\epsilon^{\mu\nu\rho\sigma} F_{\mu\nu\rho\sigma} \, |H'|^2$, then the quadratically divergent tadpole term vanishes at one-loop. In this case, the scalar fields transform under the discrete symmetry as $\phi\rightarrow -\phi$, $H\leftrightarrow H'$. Then, we can choose the same bare mass for the extra Higgs-like scalar as for the SM Higgs such that it remains decoupled during the relaxation of the Higgs mass.
For the later discussion, we assume that the effective tadpole term is chosen such that the VEV of the pseudo-scalar field is smaller than the value of the axion-like coupling $f$  in eq.~(\ref{L0}) for the valid effective theory.

\subsection{Critical four-form flux and vacuum displacement}

We find that the critical value of the flux parameter for a vanishing effective Higgs mass parameter or $v_H=0$ is given by
\bea
q_c=\frac{1}{c_H}\, \Big(M^2- c_H\mu v_\phi(q_c)\Big). \label{qcrit}
\eea
Then, solving eq.~(\ref{qcrit}) with eq.~(\ref{vev2}) for $q_c$, we get
\bea
q_c&=&\frac{\mu^2+m^2_\phi}{m^2_\phi}\,\frac{M^2}{c_H} -\mu\alpha, \\
v_\phi(q_c)&=&\alpha -\frac{\mu}{m^2_\phi} \, \frac{M^2}{c_H}\equiv v_{\phi,c}, \label{vevcrit}
\eea
and the cosmological constant at $q=q_c$ is given by
\bea
V_c&=& \Lambda+\frac{1}{2} \Big(\mu v_\phi(q_c) + q_c\Big)^2+\frac{1}{2} m^2_\phi (v_{\phi,c}-\alpha)^2 \nonumber \\
 &=&\Lambda + \frac{1}{2} \frac{m^2_\phi}{\mu^2+m^2_\phi}\, (q_c+\mu\alpha)^2.
\eea

On the other hand, electroweak symmetry is broken at $q=q_c-e$, for which
\bea
v_H(q_c-e)&=& \sqrt{\frac{|m^2_H|}{\lambda_{H,{\rm eff}}}} \equiv v, \\
v_\phi(q_c-e) &=& v_{\phi,c}    -\frac{\mu}{\mu^2+m^2_\phi} \cdot\Big( \frac{1}{2} c_H v^2 -e\Big)\equiv v_{\phi,0} \label{vevtrue}
\eea
with $|m^2_H|\equiv M^2-c_H (q_c-e+\mu v_\phi)$, and  the cosmological constant at $q=q_c-e$ is tuned to a tiny value as observed,
\bea
V_0&=& \Lambda -\frac{1}{4} \lambda_{H,{\rm eff}} v^4 +\frac{1}{2} \Big(\mu v_{\phi,0} + q_c-e\Big)^2+\frac{1}{2} m^2_\phi (v_{\phi,0}-\alpha)^2 \approx 0.
\eea

Consequently, we find that the weak scale depends on various parameters in the model, as follows,
\bea
v^2 = \frac{m^2_\phi}{\mu^2+m^2_\phi} \left(\frac{ c_H\,  e}{\lambda_{H,{\rm eff}}-  \frac{1}{2}  \frac{c^2_H\mu^2}{\mu^2+m^2_\phi}} \right).
\eea
As far as $m_\phi\sim |\mu|$, the weak scale can be obtained for the membrane charge $e$ of a similar scale, insensitive to the values of $m_\phi$ and $\mu$. But, for $m_\phi\ll |\mu|$, we can take a larger value of $e$. 
For $m_\phi\lesssim |\mu|$, which is natural for a small explicit breaking of the shift symmetry, we get the electroweak scale suppressed to
\bea
v^2\simeq \frac{m^2_\phi} {\mu^2}\,\frac{c_H e}c{\lambda_H}. \label{vev-approx}
\eea
Therefore, we can choose a larger membrane charge $e$, for instance, $\sqrt{e}\sim 1(10)\,{\rm TeV}$, for $m_\phi\sim 0.1(0.01) |\mu|$ and $c_H={\cal O}(1)$.
Moreover,  from eqs.~(\ref{vevcrit}) and (\ref{vevtrue}), after the last membrane nucleation, the pseudo-scalar VEV is shifted  by
\bea
\Delta v_\phi= v_{\phi,c}-v_{\phi,0}&=&  -\frac{\mu}{\mu^2+m^2_\phi} \cdot\Big( \frac{1}{2} c_H v^2 -e\Big) \nonumber \\
&\approx & -\frac{v^2}{\mu}\, \bigg(\frac{1}{2}c_H- \frac{\lambda_H}{c_H}\,\cdot\frac{\mu^2}{m^2_\phi}\bigg) \nonumber \\
&\approx&  \frac{\lambda_H}{c_H}\,\cdot\frac{v^2\mu}{m^2_\phi}. \label{vevshift}
\eea
where we assumed $m_\phi\lesssim |\mu|$ in the approximations. 
As a result, we can make use of the flux-induced displacement of the pseudo-scalar field for reheating, as will be discussed below.  

We remark that the pseudo-scalar VEV in the true vacuum, $v_{\phi,0}$, is model-dependent, because it depends on $\alpha$, $M^2$ and pseudo-scalar mass parameters, etc, as can be seen from eqs.~(\ref{vevcrit}) and (\ref{vevtrue}). However, we can always take $\alpha$ such that $v_{\phi,0}$ is almost zero without affecting the reheating process. In this case, we can keep the Yukawa coupling of the pseudo-scalar field to dark matter  almost CP-odd. This fact becomes important for the later discussion on the direct detection bounds for dark matter in our model.

 \subsection{Reheating}

\begin{figure}[tbp]
\centering 
\includegraphics[width=.55\textwidth]{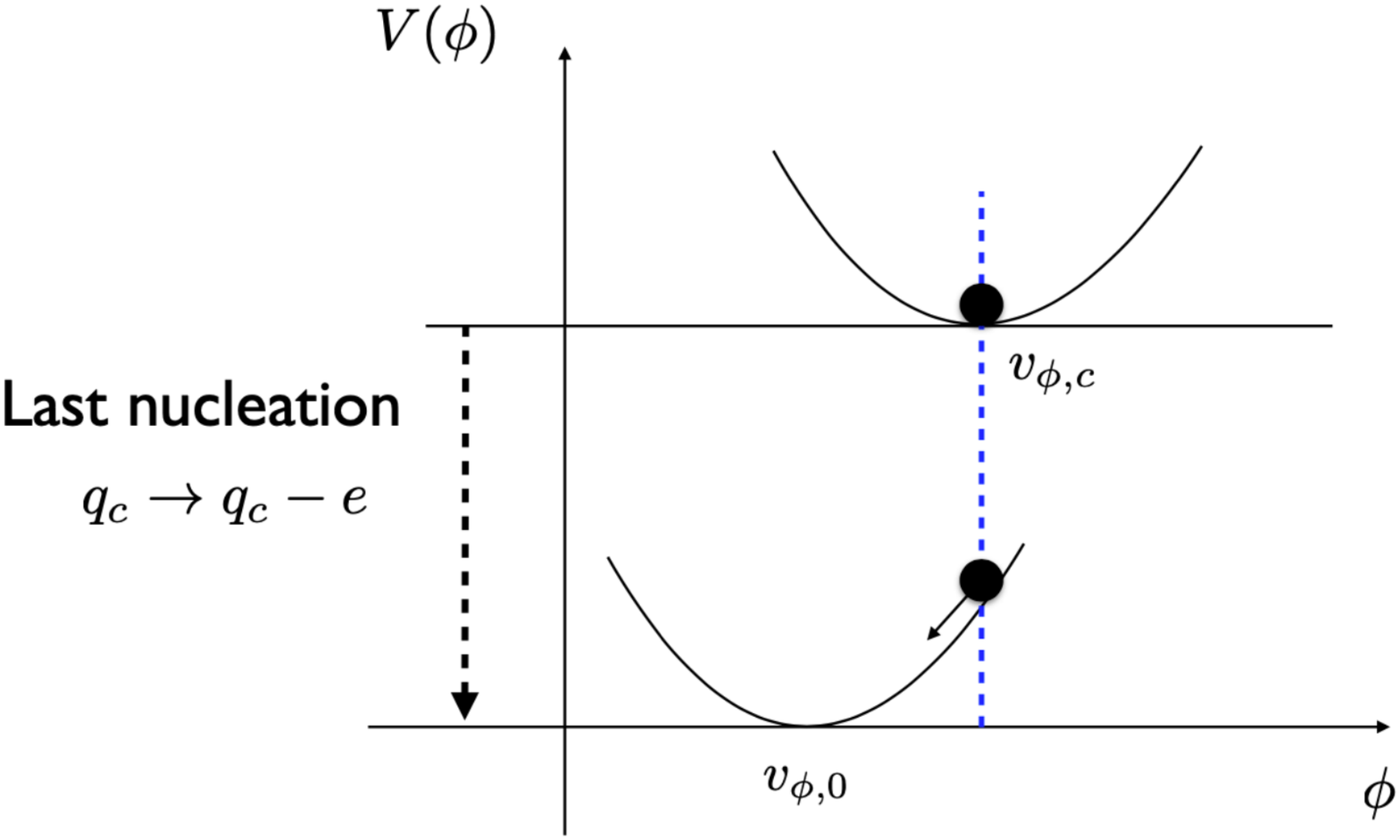} 
\caption{\label{fig:reheat}  The scalar potential for the pseudo-scalar field with the flux-dependent minima.   }
\end{figure}

Just after the last membrane nucleation, the full potential can be rewritten as
\bea
V(h,\phi)=\frac{1}{4}\lambda_{{\rm eff}}\Big( h^2-v^2\Big)^2+ \frac{1}{2}(\mu^2+m^2_\phi)\Big(\phi-v_{\phi,0}+\frac{c_H\mu}{\mu^2+m^2_\phi} (h^2-v^2)\Big)^2
\eea
where $\lambda_{{\rm eff}}=\lambda_{H,{\rm eff}}-2c^2_H\mu^2/(\mu^2+m^2_\phi)$. 
Then, setting the initial value of $\phi$ just before the last nucleation to $\phi_i=v_{\phi,c}$ and $\phi=\phi_i+\varphi$, the above potential just after the last nucleation becomes
\bea
V(h,\varphi)=\frac{1}{4}\lambda_{{\rm eff}}\Big( h^2-v^2\Big)^2+ \frac{1}{2}(\mu^2+m^2_\phi)\Big(\varphi-\Delta v_\phi+\frac{c_H\mu}{\mu^2+m^2_\phi} (h^2-v^2)\Big)^2.
\eea
Therefore, at the onset of the pseudo-scalar oscillation, with the SM Higgs frozen to $h=v$, the initial vacuum energy for reheating is given by
\bea
V_i &\equiv&  \frac{1}{2}(\mu^2+m^2_\phi) (\Delta v_\phi)^2 \nonumber \\
&=& \frac{1}{2} \frac{\mu^2}{\mu^2+m^2_\phi}\, \cdot\Big(e- \frac{1}{2} c_H v^2\Big)^2. \label{venergy-exact}
\eea
In Fig.~\ref{fig:reheat}, we depict how the minimum of the scalar potential for the pseudo-scalar changes after the last membrane nucleation and how the initial condition for reheating sets in.

We find that it is natural to take the explicit breaking term for the shift symmetry to be small, that is, $m_\phi\lesssim| \mu|$, for which the initial vacuum energy  in eq.~(\ref{venergy-exact}) is approximated to 
\bea
V_i&\simeq&  \frac{1}{2} \Big(e- \frac{1}{2} c_H v^2\Big)^2 \nonumber  \\
&\simeq & \frac{1}{2} \bigg(\frac{\lambda_H}{c_H}\,\frac{\mu^2}{m^2_\phi}-\frac{1}{2}c_H\bigg)^2 v^4, \label{venergy-approx}
\eea
almost dependently of pseudo-scalar mass parameters.
Here, we used eq.~(\ref{vev-approx}) to eliminate $e$ in the second line of eq.~(\ref{venergy-approx}).

The pseudo-scalar field starts oscillating from the shifted value, just after the end of the last membrane nucleation, as far as $m_\varphi=\sqrt{\mu^2+m^2_\phi}>H_i=\sqrt{V_i/(3M^2_P)}$, which is about $10^{-5}-10^{-1}\,{\rm eV}$ for $\sqrt{e}\sim 100\,{\rm GeV}-10\,{\rm TeV}$.
Then, the maximum temperature of the Universe in the model would be
\bea
T_{\rm max}=\left( \frac{30 V_i}{\pi^2 g_*} \right)^{1/4}\simeq 40\,{\rm GeV} \left(\frac{V^{1/4}_i}{100\,{\rm GeV}} \right)\left(\frac{100}{ g_*} \right)^{1/4} \label{Tmax}
\eea 
Thus, choosing $\sqrt{e}\sim100\,{\rm GeV}-10\,{\rm TeV}$ for $m_\phi/|\mu|\sim 0.01-1$ and $c_H={\cal O}(1)$, we get the maximum reheating temperature as
\bea
T_{\rm max}\sim 40\,{\rm GeV}-4\,{\rm TeV}. 
\eea
Therefore, the reheating temperature would be high enough for dark matter particles with mass $m_\chi<T_{\rm max}$ to thermalize, once they are produced from the decay of the pseudo-scalar field or the scattering between the SM particles.

We now discuss the reheating from the perturbative decay of the pseudo-scalar field.
From the $\varphi$ coupling to the Higgs, ${\cal L}\supset - \frac{1}{2}c_H\mu \varphi h^2$, for $m_\varphi= \sqrt{m^2_\phi+\mu^2}>2m_h$, the perturbative decay rate of the pseudo-scalar field into two Higgs bosons is given by
\bea
\Gamma(\varphi\to hh)\simeq \frac{c^2_H \mu^2}{32\pi m_\varphi} \left(1-\frac{4m^2_h}{m^2_\varphi} \right)^{1/2}.
\eea
Then, for $c_H={\cal O}(1)$ and $|\mu|\gtrsim m_\varphi\gtrsim 0.16 v$ for $\theta^2\lesssim 0.1$ to be consistent with the Higgs data, we get $\Gamma(\varphi\to hh)\sim 0.1 m_\varphi\gtrsim 0.01 v$, for which $\Gamma_2\gg H\sim \sqrt{V_i}/(\sqrt{3} M_P)$ at $T_{\rm max}$, so the reheating is instantaneous. Therefore, the reheating temperature is given by $T_{\rm max}$ as in eq.~(\ref{Tmax}).

On the other hand, if $m_\varphi<2m_h$, the perturbative decay of the pseudo-scalar field into two Higgs bosons is kinematically closed, so we need to rely on the off-shell decay processes of the Higgs bosons, such as $\varphi\to h^* h\to b{\bar b} h$ for $m_\varphi>m_h+2m_b$ and $\varphi\to h^* h^*\to {\bar b}{\bar b}bb$ for $4m_b<m_\varphi <m_h+2m_b$.
In the former case, the ratio of the corresponding decay rate to the two-body decay rate is $\Gamma_3/\Gamma_2\sim  \frac{y^2_b}{4\pi}$, and in the latter case, it is similarly given by $\Gamma_4/\Gamma_2\sim \frac{y^4_b}{(4\pi)^2}$.
Even in these cases, as far as $\Gamma_2, \Gamma_4\gg H$ at  $T_{\rm max}$, the reheating is instantaneous, so the reheating temperature is again given by eq.~(\ref{Tmax}).

\section{Flux-mediated dark matter}

We first discuss the dark matter interactions through the pseudo-scalar mediator with the four-form couplings and determine the dark matter abundance from freeze-out. Then, we consider the bounds from indirect and direct detection experiments and Higgs searches and electroweak data.

\subsection{Dark matter interactions}

From the Lagrangian for dark matter in eq.~(\ref{L0}), in the original basis with $\chi'=(\chi'_1,\chi'_2)^T$, we get the mass term shifted due to the VEV of the pseudo-scalar field by
\bea
{\cal L}_{\chi,{\rm mass}} &=& -m'_\chi {\bar \chi}' \chi' +\frac{im'_\chi v_\phi}{f} \, {\bar\chi}' \gamma^5 \chi' \nonumber \\
&=&-m'_\chi\Big(1-\frac{i v_\phi}{f} \Big) \chi^{\prime\dagger}_1 \chi'_2 -m'_\chi\Big(1+\frac{i v_\phi}{f} \Big) \chi^{\prime\dagger}_2 \chi'_1 \nonumber \\
&=& -m_\chi {\bar\chi} \chi
\eea
where 
\bea
m_\chi=m_\chi'\sqrt{1+\frac{v^2_\phi}{f^2}}=\frac{m'_\chi}{\cos\beta}, \qquad \tan\beta=\frac{v_\phi}{f},
\eea
and $\chi=(\chi_1,\chi'_2)^T$ is the redefined dark matter fermion with $\chi_1=e^{i\beta} \chi'_1$.
Moreover, in the basis of mass eigenstates for the scalar fields, we obtain the interaction terms for dark matter as follows,
\bea
{\cal L}_{\chi,{\rm int}} &=&\frac{im'_\chi }{f}\, \Big( \cos\theta\, h_1 + \sin\theta\, h_2\Big) {\bar\chi}' \gamma^5 \chi' \nonumber \\
&=& \frac{im'_\chi }{f}\, \Big( \cos\theta\, h_1 + \sin\theta\, h_2\Big) \Big(e^{i\beta} \chi^{\dagger}_1 \chi'_2 - e^{-i\beta} \chi^{\prime\dagger}_2 \chi_1 \Big) \nonumber \\
&=& \frac{im'_\chi }{f}\, e^{i\beta}\Big( \cos\theta\, h_1 + \sin\theta\, h_2\Big) {\bar\chi} P_R \chi  \nonumber \\
&&-\frac{im'_\chi }{f}\, e^{-i\beta}\Big( \cos\theta\, h_1 + \sin\theta\, h_2\Big) {\bar\chi} P_L \chi \nonumber \\
&\equiv & -\sum_{i=1,2}h_i {\bar\chi} \Big(v_{\chi,i}+i a_{\chi,i}\gamma^5 \Big)\chi \label{DMint}
\eea
where the projection operators are given by $P_L=\frac{1}{2}(1-\gamma^5)$ and $P_R=\frac{1}{2}(1+\gamma^5)$, and the CP-even and CP-odd Yukawa couplings are
\bea
v_{\chi,1} &=&\frac{m'_\chi}{f} \sin\beta\cos\theta, \qquad
a_{\chi,1}=-\frac{m'_\chi}{f} \cos\beta\cos\theta, \\
v_{\chi,2} &=&\frac{m'_\chi}{f} \sin\beta \sin\theta,  \qquad
a_{\chi,2} = -\frac{m'_\chi}{f} \cos\beta \sin\theta.
\eea
Then, a nonzero VEV of the pseudo-scalar field also gives rise to a nonzero CP-even coupling between the singlet-like scalar and dark matter. The Higgs mixing leads to the direct CP-even and CP-odd couplings between the SM-like Higgs and dark matter.

We also find that the Yukawa couplings between the SM Higgs and the SM fermions $f$ (quarks or leptons) gives rise to
\bea
{\cal L}_Y &=& -  \frac{m_f}{v}\, h \, {\bar f} f  \nonumber \\
&\equiv & - \sum_{i=1,2} v_{f,i}h_i {\bar f} f . \label{SMint} 
\eea
with
\bea
v_{f,1} = -\frac{m_f}{v}\,\sin\theta,  \qquad
v_{f,2} =  \frac{m_f}{v}\, \cos\theta.
\eea
Then, the singlet-like scalar has a CP-even coupling to the SM fermions through the Higgs mixing.
There are Higgs-like interactions between the extra scalar field and the other particles in the SM such as massive gauge bosons at tree level and massless gauge bosons at loop level \cite{axionmed}. 
We note that the pseudo-scalar couples to the SM only through the Higgs mixing, so the constraints from electric dipole moments on the axion-like scalar field do not apply in our case \cite{relaxion}.

As a result, due to the broken CP symmetry in the four-form interactions,  there exist both CP-even and CP-odd scalar interactions between scalars and the dark matter fermion, due to the Higgs mixing. But, for $v_\phi\lesssim f$ or $|\beta|\lesssim 1$, the Yukawa couplings to dark matter are like CP-odd scalar interactions, so it is possible to make the dark matter annihilation into the SM fermions to be $s$-wave. On the other hand,  the DM-nucleon scattering cross section is suppressed by the velocity of dark matter. Therefore, the DM annihilation can be relevant for indirect detection experiments, being compatible with strong direct detection bounds such as XENON1T.

We also obtain the mediator interactions from the following scalar self-interactions for pseudo-scalar and Higgs,
\bea
{\cal L}_{\rm scalar, int} = - c_H\mu \,\phi |H|^2-\lambda_{H,{\rm eff}} |H|^4, \label{sself}
\eea
The details of the scalar self-interactions in the basis of mass eigenstates are given in Appendix A.
Here, the product of four-form couplings, $\mu c_H$, is expressed in terms of the Higgs mixing angle and the scalar mass parameters from eq.~(\ref{mixing}), as follows,
\bea
c_H\mu =\frac{1}{2} (m^2_\varphi-m^2_h) \tan(2\theta)\approx (m^2_{h_1}-m^2_{h_2}) \theta
\eea
where we made an approximation for $\theta\ll 1$ in the end. 
Moreover, the effective Higgs quartic coupling $\lambda_{H,{\rm eff}}$ is approximately related to the Higgs mass parameter by
\bea
\lambda_{H,{\rm eff}}=\frac{m^2_h}{2v^2} \approx \frac{m^2_{h_2}}{2v^2}.
\eea

Furthermore, due to the Higgs mixing, we also obtain the effective interactions between scalars and massless gauge bosons in the SM, namely, photons and gluons \cite{djouadi}, respectively,
\bea
{\cal L}_{\gamma,g} &=& -\frac{\alpha_{\rm em}}{8\pi v}\,A_\gamma\, (-\sin\theta\,h_1+\cos\theta\,h_2) F_{\mu\nu} F^{\mu\nu} \nonumber \\
&&-\frac{\alpha_S}{12\pi v}\,A_g\,(-\sin\theta\,h_1+\cos\theta\,h_2) G_{\mu\nu} G^{\mu\nu} \label{masslessgauge}
\eea
where $A_\gamma, A_g$ are the loop functions, given by
\bea
A_\gamma&=& A_V(\tau_W) + N_c Q^2_t A_f (\tau_t), \\
A_g&=& \frac{3}{4} A_f(\tau_t),
\eea
with $\alpha_S=g^2_S/(4\pi)$, $\tau_W=M^2_h/(4M^2_W)$, $\tau_t=M^2_h/(4m^2_t)$, and
\bea
A_V(x) &=&-x^{-2} \Big[ 2x^{2} +3x +3(2x-1)f(x) \Big],   \\
A_f(x) &=& 2x^{-2} \Big[x+(x-1) f(x) \Big],
\eea
and
\bea
f(x) =\left\{\begin{array}{c}  {\rm arcsin}^2\sqrt{x}, \qquad\quad  x\leq 1, \vspace{0.2cm} \\  -\frac{1}{4}\bigg[\ln \frac{1+\sqrt{1-x^{-1}}}{1-\sqrt{1-x^{-1}}} -i\pi \bigg]^2, \quad x>1.  \end{array} \right.
\eea
Here, we note that the electromagnetic and strong couplings are given by $\alpha_{\rm em}(M_Z)=\frac{1}{128.9}$ and $\alpha_S(M_Z)=0.118$ at $Z$-pole, respectively, and in the limit of $\tau_{t}\ll 1$, the loop functions are approximated to $A_f(\tau_t) \rightarrow \frac{4}{3}$ and $A_g \rightarrow 1$.

Consequently, fixing $m_{h_2}=125\,{\rm GeV}$ for the mass of the SM-like Higgs, we have five independent parameters for dark matter, as follows,
\bea
m_\chi, \quad m_{h_1}, \quad  f,\,\quad \beta,\quad \theta.
\eea
Here, $\beta={\rm arctan}(v_\phi/f)$ stands for the VEV of the pseudo-scalar field, and $\theta$ is the mixing between the Higgs and pseudo-scalar fields.

\subsection{Dark matter annihilations}

Since the maximum reheating temperature is limited by about $T_{\rm max}=40\sim 4000\,{\rm GeV}$ in this model, dark matter lighter than $T_{\rm max}$ is automatically produced while being relativistic, so the freeze-out process would follow immediately for WIMP-like dark matter.

On the other hand, if dark matter is heavier than $T_{\rm max}$,  the initial dark matter abundance from thermalization is Boltzmann-suppressed by the reheating temperature. Instead, dark matter can be produced from the decay of the pseudo-scalar field if kinematically allowed and reannihilate.  In either case, the dark matter abundance is suppressed as compared to the case with $m_\chi< T_{\rm max}$, even before the freeze-out mechanism kicks in. So, in the later discussion, we focus on the case with $m_{\chi}< T_{\rm max}$ such that the freeze-out mechanism determines the dark matter abundance. 

First, dark matter can pair annihilate into a pair of the SM fermions. Then, for the non-relativistic dark matter, the corresponding annihilation cross section before thermal average is given by
\bea
(\sigma v_{\rm rel})_{\chi{\bar\chi}\to f{\bar f}}  \simeq\frac{m^2_f m^4_\chi }{8\pi v^2   f^2}\,\cos^4\beta \sin ^22\theta \bigg(\frac{1}{4m^2_\chi-m^2_{h_1}}-\frac{1}{4m^2_\chi-m^2_{h_2}}\bigg)^2  \bigg(1-\frac{m^2_f}{m^2_\chi} \bigg)^{3/2}.
\eea
Here, we ignored the velocity-dependent terms for dark matter, which are given by eq.~(\ref{ff-full}) in Appendix B. 
Then, the above channels are $s$-wave, so they are relevant for the indirect detection of dark matter from cosmic ray observations.

Moreover, for $m_\chi>m_{h_1},m_{h_2}$, dark matter can also annihilate into a pair of scalars, $h_1h_1$, $h_2 h_2$ and $h_1 h_2$. The corresponding cross sections, in the limit of a small Higgs mixing angle, are given by
\bea
&&( \sigma v_{\rm rel})_{\chi\bar{\chi}\rightarrow h_1 h_1}\simeq {m_{\chi}^2 \cos^4\beta \sqrt{1-{m_{h_1}^2\over m_{\chi}^2}}\over 128 \pi f^4 (8m_{\chi}^4 -6m_{\chi}^2 m_{h_1}^2 + m_{h_1}^4)^2(4m_{\chi}^2-m_{h_2}^2)^2}  \nonumber \\
&&\times \bigg[ 4m_{\chi}^2(4m_{\chi}^2 - m_{h_1}^2)(4m_{\chi}^2-m_{h_2}^2)\cos^2\theta\sin 2\beta - f(2m_{\chi}^2-m_{h_1}^2)\sin^2\theta   \nonumber  \\
&&\quad\times \Big\{ c_H \mu (8m_{\chi}^2+m_{h_1}^2-3m_{h_2}^2)+3(m_{h_1}^2-m_{h_2}^2)\big( c_H \mu \cos 2\theta- 2\lambda_{H,{\rm eff}}v \sin 2\theta \big)  \Big\} \bigg]^2 \nonumber \\
&&+  {m_\chi^6 \cos^4\beta \cos^4\theta \sqrt{1-{m_{h_1}^2\over m_{\chi}^2}} \, v_{\rm rel}^2 \over 384\pi f^4(m_\chi^2-m_{h_1}^2)(2m_\chi^2-m_{h_1}^2)^4 }  \Big(24m_\chi^6-60m_\chi^4 m_{h_1}^2+54m_\chi^2 m_{h_1}^4-15m_{h_1}^6 \nonumber \\
&&\hspace{0.3cm} -8(8m_\chi^6-14m_\chi^4m_{h_1}^2+7m_\chi^2 m_{h_1}^4-m_{h_1}^6)\cos2\beta \nonumber \\
&&\hspace{0.3cm} +(56m_\chi^6-100m_\chi^4m_{h_1}^2+50m_\chi^2m_{h_1}^4-9m_\chi^6)\cos4\beta   \Big),
\eea
\bea
&&( \sigma v_{\rm rel})_{\chi\bar{\chi}\rightarrow h_2 h_2}\simeq {m_{\chi}^2 \cos^4\beta \sqrt{1-{m_{h_2}^2\over m_{\chi}^2}}\over 128 \pi f^4 (8m_{\chi}^4 -6m_{\chi}^2 m_{h_2}^2 + m_{h_2}^4)^2(4m_{\chi}^2-m_{h_1}^2)^2}  \nonumber \\
&&\times \bigg[ 4m_{\chi}^2(4m_{\chi}^2 - m_{h_1}^2)(4m_{\chi}^2-m_{h_2}^2)\sin^2\theta \sin 2\beta - f(2m_{\chi}^2-m_{h_2}^2)\cos^2\theta \nonumber \\
&&\quad\times\Big\{ c_H\mu (8m_{\chi}^2+m_{h_2}^2-3m_{h_1}^2) +3(m_{h_1}^2-m_{h_2}^2)\big( c_H \mu \cos 2\theta-2\lambda_{H,{\rm eff}}v \sin 2\theta \big)  \Big\} \bigg]^2, 
\eea
\bea
&&(\sigma v_{\rm rel})_{\chi{\bar\chi}\to h_1 h_2} \simeq {m_{\chi}^2\cos^4 \beta \sin^2 2\theta \sqrt{1-{m_{h_1}^2+m_{h_2}^2\over 2m_{\chi}^2}+{(m_{h_1}^2-m_{h_2}^2)^2\over 16 m_{\chi}^4}} \over 256 \pi f^4 (4m_\chi^2-m_{h_1}^2)^2(4m_\chi^2-m_{h_2}^2)^2(4m_\chi^2-m_{h_1}^2-m_{h_2}^2)^2} \nonumber \\
&&\times \bigg[8m_\chi^2 (4m_\chi^2-m_{h_1}^2)(4m_\chi^2-m_{h_2}^2)\sin 2\beta  +f(4m_\chi^2-m_{h_1}^2-m_{h_2}^2)  \\
&&\quad\times \Big\{ c_H \mu (8m_\chi^2-m_{h_1}^2-m_{h_2}^2)+3(m_{h_1}^2-m_{h_2}^2) \big( c_H \mu \cos 2\theta-2\lambda_{H,{\rm eff}}v \sin 2\theta \big)  \Big\}  \bigg]^2.
\eea
For a sizable Higgs mixing, the correction terms for $\chi{\bar \chi}\rightarrow h_1 h_1$  are given in eq.~(\ref{h1h1full}) in Appendix B.

For a small $|\sin\beta|$, which is favored for direct detection as will be discussed in the later  subsection, the dark matter annihilation into a pair of singlet-like scalars ($h_1 h_1$) has the s-wave contribution suppressed while the p-wave contribution unsuppressed during freeze-out even for small Higgs mixing. Thus, the $h_1 h_1$ channel is important for determining the correct relic density, as will be shown in the later subsection.  
On the other hand, the dark matter annihilation into a pair of SM-like Higgs bosons ($h_2 h_2$) is s-wave dominant, but it is suppressed  because  it depends on the Higgs mixing angle and the product of the four-form flux coupling, $c_H\mu$, which is bounded by a small Higgs mixing angle in eq.~(\ref{mixing}). For relatively light dark matter below the electroweak scale, the $h_2 h_2$ channel is kinematically closed. 
Finally, the dark matter annihilation into one singlet-like scalar and one SM-like Higgs scalar ($h_1 h_2$) is also suppressed by the Higgs mixing angle, but it is s-wave. 

For $m_\chi>m_W, m_Z$, we also need to consider the dark matter annihilations into a pair of massive gauge bosons in the SM, with the cross sections,
\bea
(\sigma v_{\rm rel})_{\chi\bar{\chi}\rightarrow W^+W^-}
&\simeq & {m_{\chi}^2 \over 16 \pi f^2 v^2 } \, \Big(4m_{\chi}^4-4m_W^2 m_{\chi}^2 +3m_W^4\Big)\nonumber \\
&&\quad \times\cos^4\beta\sin^2 2\theta   \bigg(\frac{1}{4m^2_\chi-m^2_{h_1}}-\frac{1}{4m^2_\chi-m^2_{h_2}} \bigg)^2 \sqrt{1-{m_W^2\over m_{\chi}^2}}
\eea
and
\bea
(\sigma v_{\rm rel})_{\chi\bar{\chi}\rightarrow ZZ}&\simeq &  { m_{\chi}^2 \over 32\pi  f^2 v^2 } \, \Big(4m_{\chi}^4-4m_Z^2 m_{\chi}^2 +3m_Z^4\Big)\nonumber \\
&&\quad \times\cos^4\beta\sin^2 2\theta   \bigg(\frac{1}{4m^2_\chi-m^2_{h_1}}-\frac{1}{4m^2_\chi-m^2_{h_2}} \bigg)^2 \sqrt{1-{m_Z^2\over m_{\chi}^2}}.
\eea
Thus, we find that the above annihilation channels into $WW, ZZ$ are suppressed by the Higgs mixing angle.

Finally, due to the Higgs mixing, the pseudo-scalar has the effective couplings to photons and gluons in eq.~(\ref{masslessgauge}), just like the SM Higgs. Then, we also get the cross sections for dark matter annihilations into a pair of photons or gluons by
\bea
(\sigma v_{\rm rel})_{\chi\bar{\chi}\rightarrow \gamma\gamma} &\simeq & \Big(\frac{\alpha_{\rm em}}{8\pi v} \Big)^2|A_\gamma|^2 \Big(\frac{m^6_\chi}{\pi f^2}\Big) \cos^4\beta \sin^22\theta \bigg(\frac{1}{4m^2_\chi-m^2_{h_1}}-\frac{1}{4m^2_\chi-m^2_{h_2}} \bigg)^2, \\
( \sigma v_{\rm rel})_{\chi\bar{\chi}\rightarrow gg} &\simeq &\Big(\frac{\alpha_S}{12\pi v} \Big)^2|A_g|^2 \Big(\frac{8m^6_\chi}{\pi f^2}\Big) \cos^4\beta \sin^22\theta \bigg(\frac{1}{4m^2_\chi-m^2_{h_1}}-\frac{1}{4m^2_\chi-m^2_{h_2}} \bigg)^2.
\eea
Consequently, we can see that  the above annihilation channels into $\gamma\gamma, gg$ are suppressed by the Higgs mixing angle as well as the loop factors.

\subsection{Indirect detection}

As dark matter can annihilate directly into $b{\bar b}$ or a pair of the SM particles through the pseudo-scalar or Higgs boson without velocity suppression,  indirect detection experiments and Cosmic Microwave Background measurements \cite{indirect,LQ} can constrain dark matter with weak-scale masses.
There are gamma-ray limits on the dark matter annihilation from Fermi-LAT dwarf galaxies \cite{fermi-lat} and HESS gamma-rays \cite{hess} and AMS-02 antiprotons \cite{ams}, constraining the model.
We can also discuss the region for explaining the gamma-ray excess at the galactic center and the cosmic ray anti-proton excess with the dark matter annihilation in our model \cite{GC}.  

Regarding the gamma-ray excess at the galactic center (GC), we remark that fermion dark matter with a mass in the range of $m_\chi=40-70$ GeV is needed for the case of annihilations to $b{\bar b}$ with about the thermal cross section, $\langle \sigma v_{\rm rel}\rangle\sim 10^{-26}\,{\rm cm}^3/s$ \cite{GC}. Moreover, the same dark matter annihilation into $b{\bar b}$  can account for the antiproton excess measured by AMS-02 for DM masses in the range of $46-94$ GeV \cite{GC}. Then, we can take the annihilation cross section into $b{\bar b}$ to be $\langle\sigma v_{\rm rel}\rangle = (0.6-7)\times10^{-26}\, {\rm cm^3/s}$ for the gamma-ray excess and $\langle\sigma v_{\rm rel}\rangle = (0.3-20)\times10{-26}\, {\rm cm^3/s}$ for the antiproton excess \cite{GC}.  Although the galactic center excess and the AMS-02 anti-proton excess are not conclusive at the moment, we indicate the region of the parameter space later favored to explain those excesses in our model for a future reference. 

It is remarkable that there might be also interesting signatures for indirect detection from the s-wave contribution of the $h_1 h_1$ channel due to the cascade decays of the scalars such as $b{\bar b}b{\bar b}$, $b{\bar b}\tau{\bar \tau}$ or $\tau{\bar \tau}\tau{\bar \tau}$ \cite{cascade,multistep}. Moreover,  there could be similar indirect signatures from the $h_1 h_2$ channel if kinematically allowed. In the case of multi-step annihilations of dark matter, the peak of the resulting gamma-ray spectrum is shifted toward the low energy and the annihilation cross section required for fitting the GC gamma-ray excess increases, as compared to the case of direct annihilation of dark matter into $b{\bar b}$ \cite{multistep}. Since the annihilation cross sections for the s-wave part of the $h_1 h_1$ and $h_1 h_2$ are suppressed by the Higgs mixing angle or $\sin\beta$, it is sufficient for us to focus on the direct annihilation of dark matter into $b{\bar b}$  in the later discussion on the bound from Fermi-LAT or the excess from AMS-02. But, it would be worthwhile to perform a dedicated analysis for the cascade decays of the singlet scalars for indirect detection.

\subsection{Direct detection}

Due to the flux-induced Higgs portal coupling for the pseudo-scalar field, ${\cal L}\supset - \frac{1}{2}c_H\mu \varphi h^2$, with a Higgs mixing, as discussed previously,  the pseudo-scalar field can communicate between dark matter and the SM, with the same four-form flux couplings.  In this case, the direct detection cross section for fermion dark matter is suppressed by the momentum transfer between dark matter and nucleon, due to the chiral operator $\gamma^5$ in the mediator coupling for dark matter \cite{axionmed}. This interesting behavior is due to the fact that the four-form couplings to both pseudo-scalar and Higgs fields exist, violating the CP symmetry.

After integrating out the pseudo-scalar and Higgs bosons, from eqs.~(\ref{DMint}) and (\ref{SMint}), we get the effective interactions between dark matter and the SM fermions, as follows,
\bea
{\cal L}_{\rm DD}&=&  \sum_f \sum_{i=1,2} \frac{ v_{f,i}}{m^2_{h_i}}\, \Big[{\bar\chi} \big(v_{\chi,i}+i a_{\chi,i}\gamma^5 \big)\chi\Big] {\bar f} f \nonumber \\
&=&\sum_f \lambda_f \sum_{i=1,2} \frac{ {\tilde v}_{i}}{m^2_{h_i}}\, \Big[{\bar\chi} \big(v_{\chi,i}+i a_{\chi,i}\gamma^5 \big)\chi\Big] {\bar f} f
\eea
where $v_{f,i}={\tilde v}_i \lambda_f$ with ${\tilde v}_1=-\sin\theta(q)$, ${\tilde v}_2=\cos\theta(q)$ and $\lambda_f=m_f/v$.

Then, for the direct detection of dark matter, we can approximate the cross section for the elastic scattering between dark matter and nucleus to
\bea
\sigma_{\chi-N} 
\simeq \frac{\mu^2_{\chi N} m^2_\chi}{4\pi v^2 f^2 A^2}\,(\sin2\theta)^2\cos^2\beta \sin^2\beta\Big(\frac{1}{m^2_{h_1}}-\frac{1}{m^2_{h_2}} \Big)^2 \Big(Z f_p + (A-Z) f_n\Big)^2 
\eea
where $Z, A-Z$ are the numbers of protons and neutrons in the detector nucleus, $\mu_{\chi N}=m_\chi m_N/(m_\chi+m_N)$ is the reduced mass for the system of dark matter and nucleus, and
\bea
f_{p,n}=m_{p,n}\bigg(\sum_{q=u,d,s}  f^{p,n}_{Tq} +\frac{2}{9}f^{p,n}_{TG} \bigg)
\eea
with $f^{p,n}_{TG}=1-\sum_{q=u,d,s}   f^{p,n}_{Tq} $.  Here,  $f^N_{Tq}$ is the mass fraction of quark $q$ inside the nucleon $N$, defined by $\langle N|m_q {\bar q}q |N\rangle= m_N f^N_{Tq}$, and $f^{N}_{TG}$ is the mass fraction of gluon $G$  the nucleon $N$, due to heavy quarks. The numerical values are given by $f^p_{T_u}=0.023$, $f^p_{T_d}=0.032$ and $f^p_{T_s}=0.020$ for a proton and $f^n_{T_u}=0.017$, $f^n_{T_d}=0.041$ and $f^n_{T_s}=0.020$ for a neutron \cite{hisano}.
Therefore, we find that as $|\sin\beta|$ decreases,  the elastic scattering cross section between dark matter and nucleus gets an extra suppression in addition to the Higgs mixing angle. 

On the other hand, for generality, we also present the elastic scattering cross section between dark matter and electron as
\bea
\sigma_{\chi-e} \simeq   \frac{\mu^2_{\chi e}m^2_e m^2_\chi}{4\pi v^2f^2} \,(\sin2\theta)^2\cos^2\beta \sin^2\beta\Big(\frac{1}{m^2_{h_1}}-\frac{1}{m^2_{h_2}} \Big)^2
\eea
where $\mu_{\chi e}=m_\chi m_e/(m_\chi+m_e)$. The above scattering cross section for electron is again suppressed for a small $\sin\beta$.  We have not considered the details of the current bounds on the DM-electron scattering cross section in this work, because we focused on the WIMP case. However, the DM-electron scattering is relevant for detecting light dark matter with sub-GeV mass \cite{subGeV} or exothermic dark matter in XENON1T \cite{exo}.

We remark that in the case that the direct detection cross section for dark matter has a chirality suppression at tree level as discussed above, the effective interactions between dark matter and nucleus (or electron) are subject to loop corrections with two pseudo-scalar exchanges \cite{loops}, which could be important for a sizable $m_\chi/f$ and a light pseudo-scalar field.
But, the full discussion on the loop corrections including two-loop diagrams for gluon effective interactions in our case is beyond the scope of our work.

\subsection{Constraints from Higgs and electroweak data}

For $m_\chi<m_{h_2}/2$, the SM-like  Higgs can decay into a pair of dark matter fermions.
Then, the corresponding partial decay rate for $h_2\to \chi{\bar\chi}$ is given by
\bea
\Gamma(h_2\to  \chi{\bar\chi})
= \frac{m^2_\chi m_{h_2}}{8\pi f^2} \,(\sin\theta)^2\cos^2\beta \bigg[\sin^2\beta\, \bigg(1-\frac{4m^2_\chi}{m^2_{h_2}}\bigg)+\cos^2\beta  \bigg]  \bigg(1-\frac{4m^2_\chi}{m^2_{h_2}} \bigg)^{1/2}. \label{Hinv}
\eea
Then, for a nonzero Higgs mixing angle, the branching ratio of Higgs invisible decay is given by
\bea
{\rm BR}_{\rm inv} =\frac{\Gamma(h_2\to  \chi{\bar\chi})}{\Gamma_{\rm tot}} 
\eea
where $\Gamma_{\rm tot}=\cos^2\theta\,\Gamma_{\rm SM} +\Gamma(h_2\to  \chi{\bar\chi}) $ with the total decay rate of the  SM Higgs,  $\Gamma_{\rm SM}=4.2\,{\rm MeV}$, for $m_{h_2}=125\,{\rm GeV}$.
The  previous limit in 2016 on the branching ratio of Higgs invisible decay  is  ${\rm BR}_{\rm inv}<0.19$ at $90\%$ C.L.\cite{Hinv-published}, and it has been updated recently to ${\rm BR}_{\rm inv}<0.11$ at $95\%$ C.L. \cite{Hinv-recent}

Moreover, for $m_{h_1}<m_{h_2}/2$, the SM-like  Higgs can also decay into a pair of dark Higgs bosons. Then, the corresponding partial decay rate for $h_2\to h_1 h_1$ is given by
\bea
\Gamma(h_2\to h_1 h_1) 
=  \frac{ \sin^2\theta}{32\pi m_{h_2}}\,\Big[\mu c_H (\sin^2\theta-2\cos^2\theta) +6\lambda_{H,{\rm eff}}v\,\cos\theta\sin\theta\Big]^2\, \bigg(1-\frac{4m^2_{h_1}}{m^2_{h_2}} \bigg)^{1/2},
\eea
which is additive to the total decay rate of the SM Higgs.
On the other hand, for $m_{h_2}<m_{h_1}/2$, the single-like scalar can decay into a pair of the SM-like Higgs bosons, with the partial decay rate,
\bea
\Gamma(h_1\to h_2 h_2) 
=  \frac{ \cos^2\theta}{32\pi m_{h_1}}\,\Big[\mu c_H (\cos^2\theta-2\sin^2\theta) -6\lambda_{H,{\rm eff}}v\,\cos\theta\sin\theta\Big]^2\, \bigg(1-\frac{4m^2_{h_2}}{m^2_{h_1}} \bigg)^{1/2}.
\eea

We remark that the Higgs mixing gives rise to the modified Higgs production rate and the new production of the singlet-like scalar at the LHC and the modified partial decay rates of Higgs visible decay modes. 

First, the production cross section for the SM-like Higgs, for instance, the gluon fusion, and the decay rates of Higgs visible decay modes are universally suppressed by $\cos^2\theta$. If extra Higgs decays are absent or ignorable, the branching ratios of the Higgs boson are almost the same as in the SM.
In this case, from the Higgs data at the LHC, the Higgs mixing angle would be constrained to be $|\sin\theta|\lesssim 0.3$, provided that the experimental uncertainties are within $10\%$ \cite{pdg-higgs}. 

On the other hand, the singlet-like scalar can be produced at colliders similarly as for the SM Higgs boson, except that the corresponding cross section and the decay modes of the singlet-like scalar are universally suppressed by $\sin^2\theta$ as compared to those for the SM Higgs and the decay branching fractions depend on the mass of the singlet-like scalar. 
Therefore, the singlet-like scalar can be constrained by LEP, Tevatron and electroweak precision data \cite{lebedev}  and it has been also searched for at the LHC. 

For $m_{h_1}<114\,{\rm GeV}$, the LEP search with $b{\bar b}$ decay mode constrains $\sin^2\theta<\zeta^2(m_{h_1})$ with $\log_{10} \zeta^2(m)\simeq m/(60\,{\rm GeV})-2.3$ \cite{lep,lebedev}.  For instance, for $m_{h_1}=50 (70)\,{\rm GeV}$, we require $\sin^2\theta<0.034(0.074)$. Secondly, the $\rho$-parameter is corrected due to the Higgs mixing angle \cite{lebedev}, as follows,
\bea
\Delta\rho&=& \frac{3G_F}{8\sqrt{2} \pi^2}\, \bigg[ \sin^2\theta \bigg(m^2_W\ln \frac{m^2_{h_1}}{m^2_W}-m^2_Z\ln \frac{m^2_{h_1}}{m^2_W} \bigg) \nonumber \\
&&\quad+\cos^2\theta \bigg(m^2_W\ln \frac{m^2_{h_2}}{m^2_W}-m^2_Z\ln \frac{m^2_{h_2}}{m^2_W} \bigg)  \bigg].
\eea
The global fit in PDG data \cite{pdg-higgs} shows $\Delta\rho=(3.9\pm 1.9)\times 10^{-4}$, which is $2\sigma$ above the SM expectation $\rho=1$. Therefore, such a deviation would indicate that
\bea
\ln 41 (35)<\sin^2\theta \ln m_{h_1} +\cos^2\theta \ln m_{h_2}<\ln  80 (94)
\eea
at $2\sigma (3\sigma)$ where the masses are measured in GeV. For instance, we would need $\sin^2\theta>0.48 (0.31)$  for $m_{h_1}=50\,{\rm GeV}$ and $\sin^2\theta>0.76(0.49)$ for $m_{h_1}=70\,{\rm GeV}$. However, the results are not consistent with the LEP limit on the Higgs mixing angle. Therefore, we only impose the LEP limit on the Higgs mixing angle for  $m_{h_1}<114\,{\rm GeV}$  in our model. The LHC searches become important for heavy singlet-like scalars through $ZZ, h_2h_2$ decay modes, constraining the Higgs mixing angle at the level of $\sin\theta\simeq 0.3$ at best \cite{lhc}.

\subsection{Combined constraints}

We impose various constraints discussed in the previous subsections on the parameter space in our model.

\begin{figure}[tbp]
\centering 
\includegraphics[width=.45\textwidth]{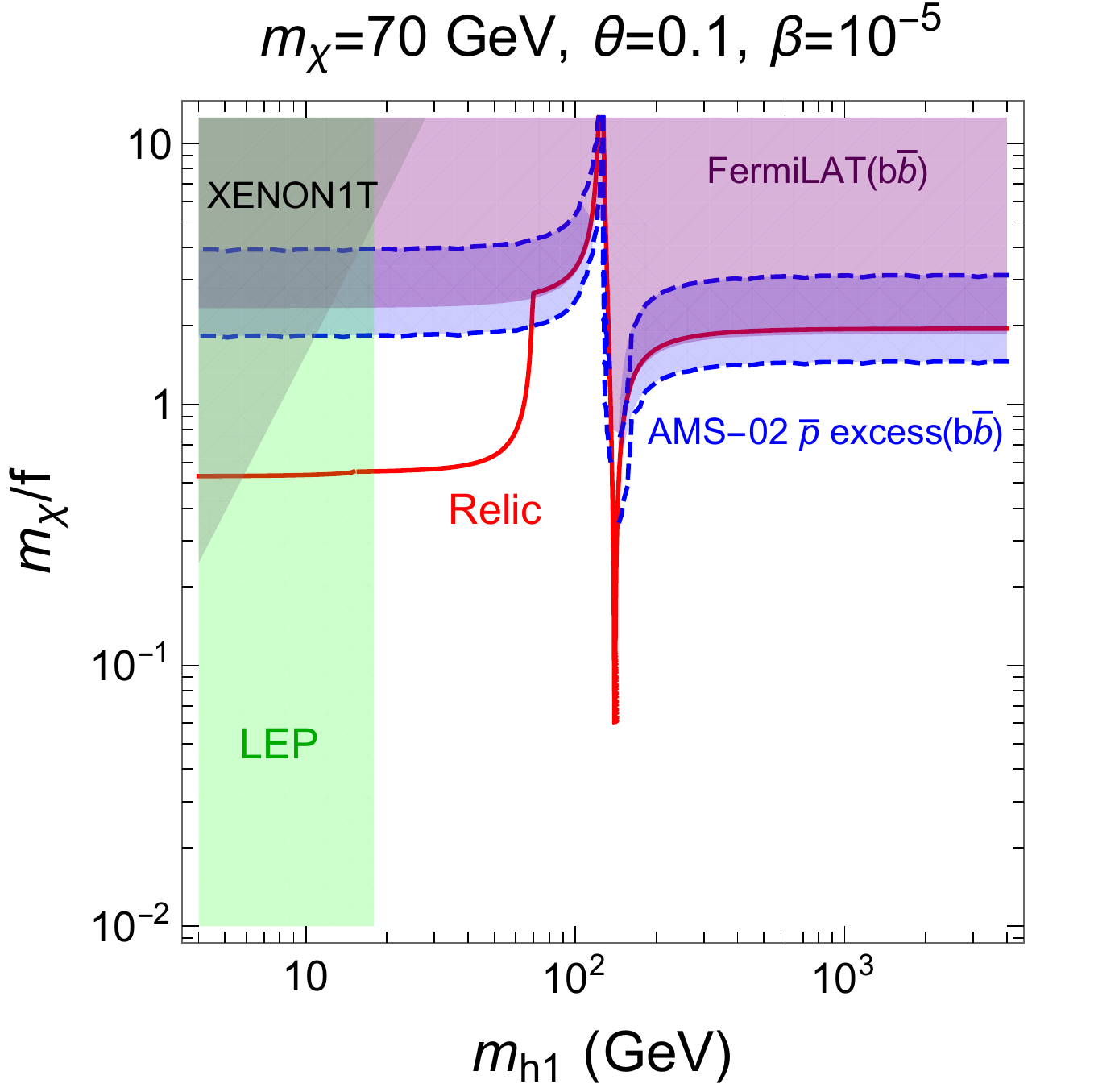} \,\,
\includegraphics[width=.45\textwidth]{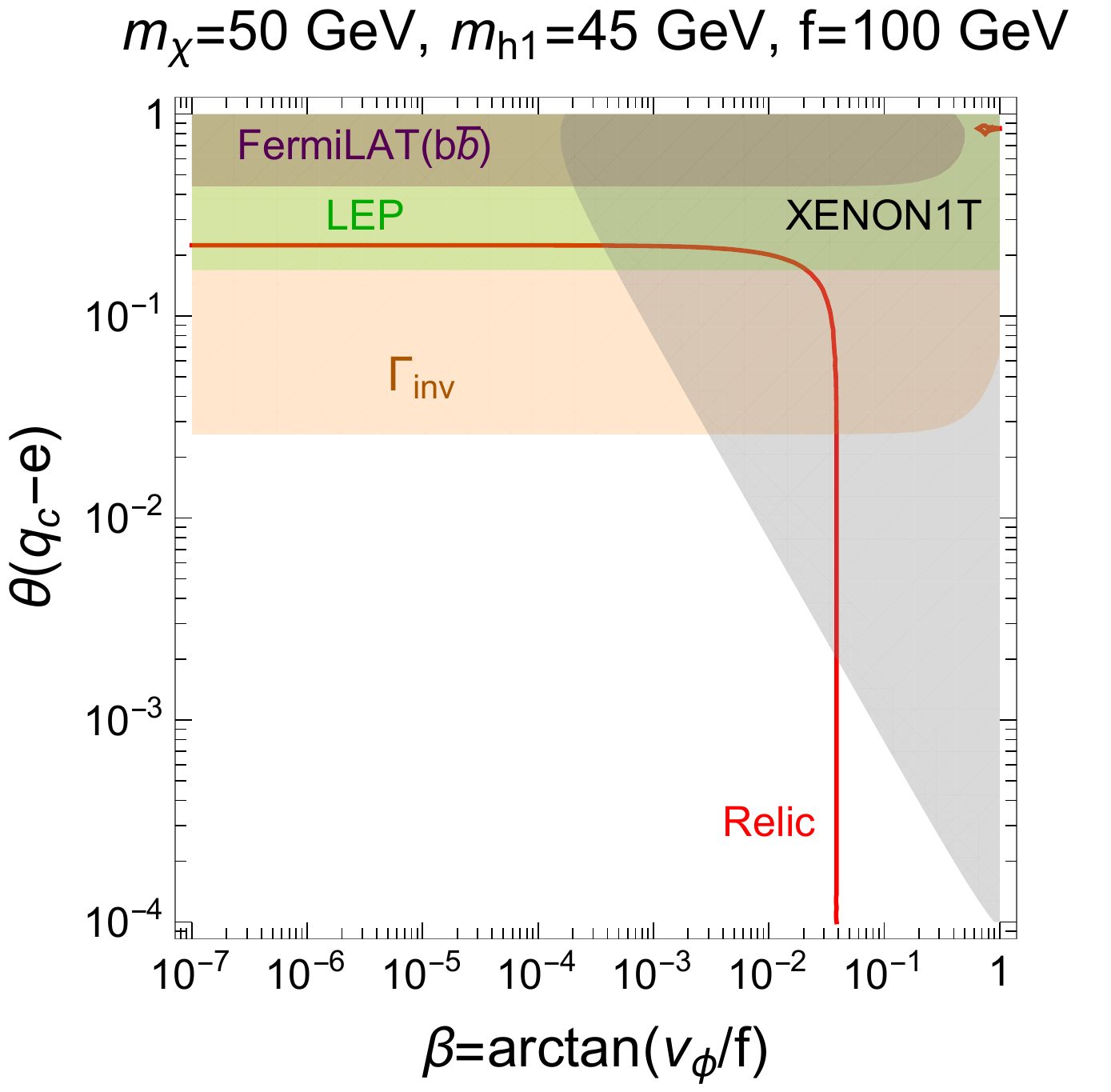} 
\caption{\label{fig:a1}  (Left) Parameter space for $m_\chi/f$ vs the singlet-like scalar mass.  We chose $m_\chi=70\,{\rm GeV}$, $\theta=0.1$ and $\beta=10^{-5}$. (Right) Parameter space for the Higgs mixing angle, $\theta(q_c-e)$ vs the pseudo-scalar VEV, $\beta={\rm arctan}(v_\phi/f)$. We chose $m_\chi=50\,{\rm GeV}$, $m_{h_1}=45\,{\rm GeV}$, $f=100\,{\rm GeV}$. The relic density is saturated along the red line. The gray and green regions are excluded by XENON1T and LEP, respectively. Purple region is disfavored by diffuse gamma-rays from Fermi-LAT dwarf galaxies ($b{\bar b}$). We also show the blue region favored by the AMS-02 anti-proton excess on left and the orange region disfavored by the bound from the Higgs invisible decay on right.  }
\end{figure}

First, in Fig.~\ref{fig:a1}, we depict the parameter space for $m_\chi/f$ vs the singlet-like scalar mass $m_{h_1}$ on left, and the parameter space for the Higgs mixing angle, $\theta$, at the relaxation of Higgs mass, vs the pseudo-scalar VEV,  parametrized by $\beta={\rm arctan}(v_\phi/f)$. The correct relic density is satisfied along the red line. For $m_\chi>m_{h_1}$, the dark matter annihilation into a pair of singlet-like scalars ($h_1h_1$) is a dominant channel for determining the relic density,  because the corresponding annihilation cross section is p-wave but unsuppressed by either the Higgs mixing angle and $\sin\beta$. On the other hand, for $m_\chi<m_{h_1}$, we need a larger dark matter coupling, $m_\chi/f$, for a fixed Higgs mixing angle, as shown in the left plot of   Fig.~\ref{fig:a1}. 

We also show in Fig.~\ref{fig:a1} that the gray and green regions are excluded by the direct detection from XENON1T and the Higgs-like scalar search with $b{\bar b}$ mode at LEP. The purple and orange regions are disfavored by the bounds from Fermi-LAT dwarf galaxies (for $b{\bar b}$ annihilation channel) discussed in the previous subsection and the Higgs invisible decay that will be discussed in the next subsection, respectively. 
We also indicated the blue region favored to explain the AMS-02 anti-proton excess \cite{GC}, but there is no consistent region to explain the Fermi-LAT gamma-ray excess at the galactic center \cite{GC}, because of the bounds from Fermi-LAT dwarf galaxies.

We took two benchmark scenarios:  the case with $m_\chi=70\,{\rm GeV}$, $\theta=0.1$ and $\beta=10^{-5}$ on the left plot in Fig.~\ref{fig:a1}, and the case with  $m_\chi=50\,{\rm GeV}$, $m_{h_1}=45\,{\rm GeV}$ and $f=100\,{\rm GeV}$ on the right plot in Fig.~\ref{fig:a1}. 
In the latter case, the Higgs invisible decay is open so the parameter space with a sizable mixing angle is disfavored. On the other hand, in the former case, there is no Higgs invisible decay, so there is a parameter space with a sizable mixing where the LEP bound for light singlet-like scalars with $m_{h_1}\lesssim 114\,{\rm GeV}$ as well as the LHC limits, $\theta\lesssim 0.3$, for heavy singlet-like scalars from the $ZZ, h_2h_2$ decay modes \cite{lhc}, are satisfied.

We find that a sizable Higgs mixing angle is constrained by the LHC data from the Higgs visible and invisible decays as well as the bounds from Fermi-LAT dwarf galaxies.  As shown on the right plot in Fig.~\ref{fig:a1}, the XENON1T bounds become more important than the bound from the Higgs invisible decay for $\beta\gtrsim 10^{-3}$.  The region with a sizable Higgs mixing angle can be searched for by indirect detection experiments, such as gamma-ray and anti-proton searches in Fermi-LAT and AMS-02 experiments, respectively. Indeed, the anti-proton excess from AMS-02 could be explained in the region of the saturated relic density, as shown on the left plot in Fig.~\ref{fig:a1}.

\begin{figure}[tbp]
\centering 
\includegraphics[width=.45\textwidth]{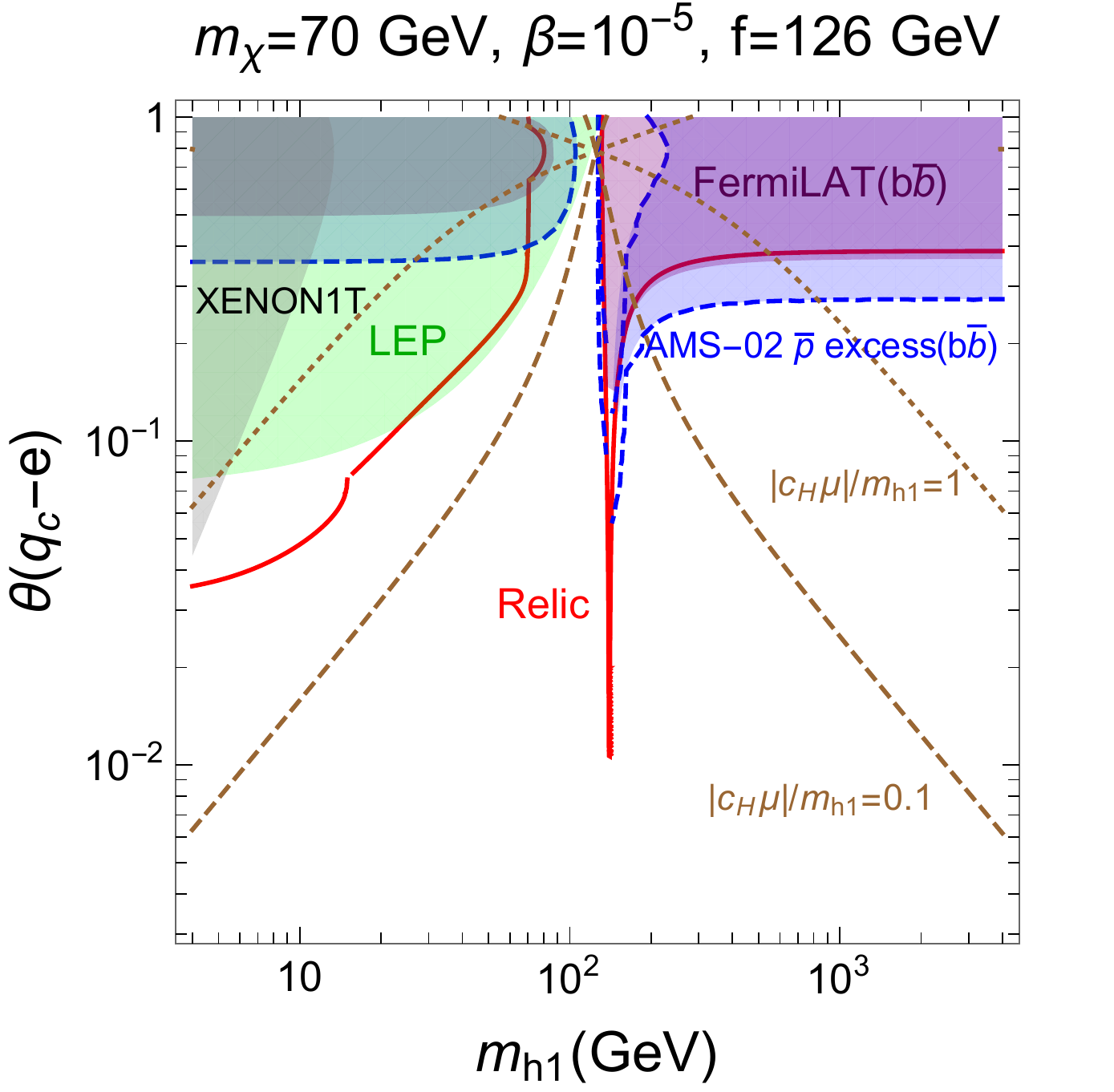} \,\,
\includegraphics[width=.45\textwidth]{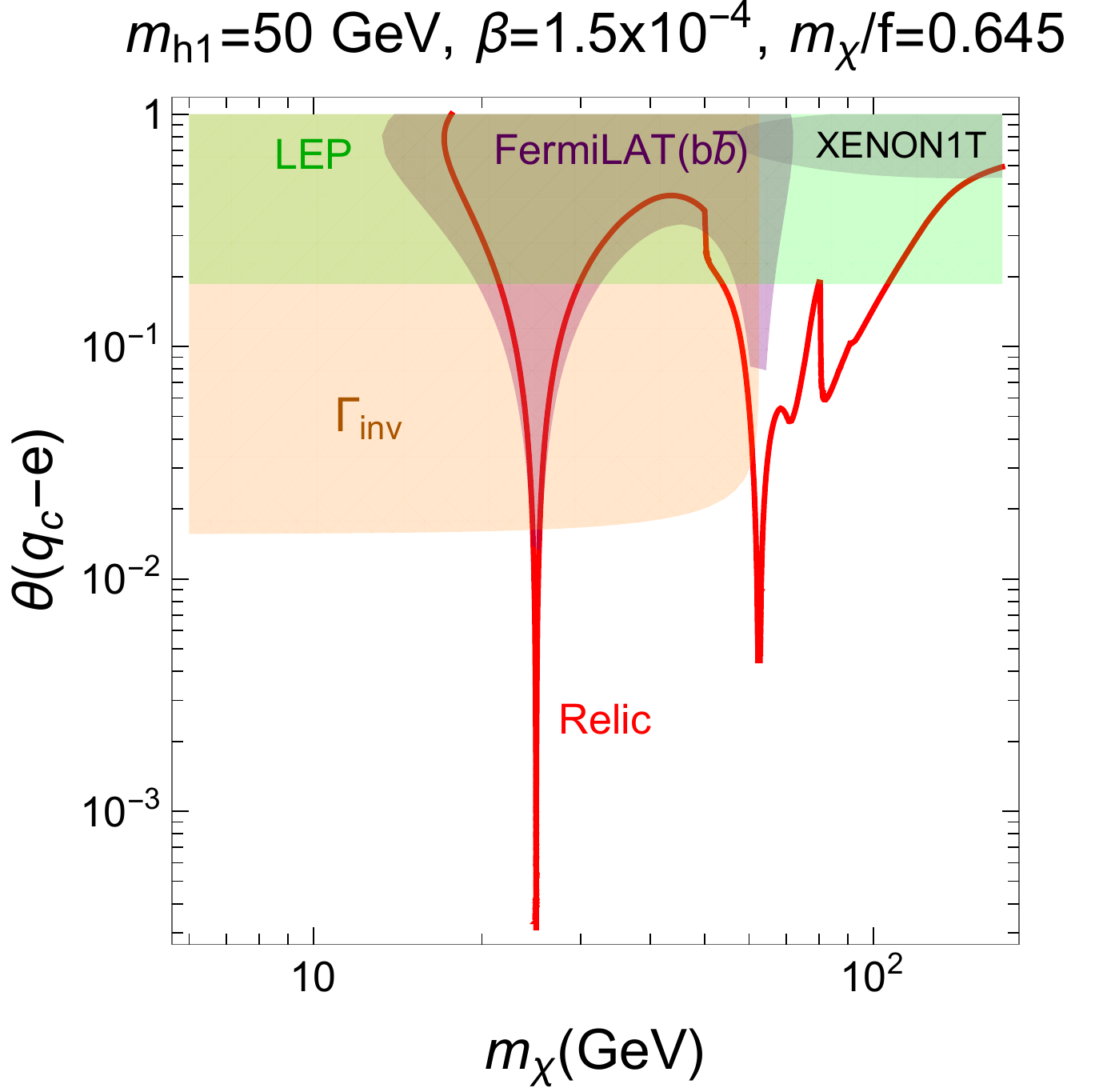}
\caption{\label{fig:a2} Parameter space for the Higgs mixing angle, $\theta(q_c-e)$, vs the singlet-like scalar mass $m_{h_1}$ on left (dark matter mass $m_\chi$ on right). We took $m_\chi=70\,{\rm GeV}$, $\beta=10^{-5}$ and $f=126\,{\rm GeV}$ on left and  $m_{h_1}=50\,{\rm GeV}$, $\beta=1.5\times 10^{-4}$ and $m_\chi/f=0.645$ on right. We also drew the contours for $|c_H \mu|/m_{h_1}=1, 0.1$ in dotted and dashed brown lines on left.  The color notations are the same as in Fig.~\ref{fig:a1}. }
\end{figure}

Next, in Fig.~\ref{fig:a2}, we draw the parameter space for the Higgs mixing angle, $\theta$,  at the relaxation of Higgs mass, vs the dark Higgs mass $m_{h_1}$ on left (the dark matter mass $m_\chi$ on right).   The colored notations for various regions are the same as in Fig.~\ref{fig:a1}. We took $m_\chi=70\,{\rm GeV}$, $\beta=10^{-5}$ and $f=126\,{\rm GeV}$ on the left plot and  $m_{h_1}=50\,{\rm GeV}$, $\beta=1.5\times 10^{-4}$ and $m_\chi/f=0.645$ on the right plot. Thus, for both cases, the singlet-like scalar coupling to dark matter is almost CP-odd, so the XENON1T limit constrains only a small region of the parameter space. 

The LEP limit excludes the region with a sizable mixing angle, for instance, the region with $\theta\gtrsim 0.2$ for $m_{h_1}=50\,{\rm GeV}$ on the right and up to $\theta\gtrsim 0.08$ for $m_{h_1}\gtrsim 4\,{\rm GeV}$ on the left in Fig.~\ref{fig:a2}. On the other hand, there is a viable region in blue with a relatively heavy singlet-like scalar on the left plot in Fig.~\ref{fig:a2} for explaining the AMS-02 anti-proton excess, whereas the purple region is disfavored by the limits from gamma-ray searches with Fermi-LAT dwarf galaxies and the LHC searches for Higgs-like scalars. We need to keep in mind that the Higgs mixing angle is constrained to  $\theta\lesssim 0.3$ for heavy singlet-like scalars from the $ZZ, h_2h_2$ decay modes at the LHC \cite{lhc}, although not shown in Fig.~\ref{fig:a2}. We also drew the contours on the left plot with $|c_H\mu|/m_{h_1}=1,0.1$ for four-form couplings in dotted and dashed brown lines, respectively. 
Noting that $m_{h_1}\simeq |\mu|$ for $m_\phi\lesssim \mu$ as well as perturbativity $|c_H|\lesssim 1$ lead to $|c_H\mu|/m_{h_1}\lesssim 1$. Thus,   from the Higgs mixing angle in eq.~(\ref{mixing}) with $m_{h_1}=50 (30)\,{\rm GeV}$, perturbativity sets $|\theta|\lesssim 0.54 (0.39)$.

The correct relic density  can be satisfied along the red line when the Higgs mixing and the dark matter coupling are sizable, even away from the resonance regions with $m_{h_1}\lesssim m_\chi$, as shown in both plots in Fig.~\ref{fig:a2}. As we have already discussed in connection to Fig.~\ref{fig:a1},  our results indicate clearly that the dark matter annihilation into a pair of singlet-like scalars ($h_1h_1$) is crucial for determining the correct relic density for a smaller Higgs mixing angle, as shown in red lines to the left region on the left plot and to the right region on the right plot Fig.~\ref{fig:a2}.
We note that the region with $m_\chi\lesssim m_{h_2}/2$ is strongly constrained by the limit from the Higgs invisible decay, except the resonance regions with $m_\chi\sim 2m_{h_1}$ or $m_\chi\sim 2m_{h_2}$, as shown on the right plot in Fig.~\ref{fig:a2}.
The resonance locations for the $b{\bar b}$ channels are velocity-dependent, so those in galaxies at present are at lower resonance masses as compared to those at freeze-out, due to the suppressed velocity of dark matter, $v\sim 10^{-3}$ or less. Therefore, if the resonant enhancement for the $b{\bar b}$ channel occurs during freeze-out, we could avoid the strong bounds from Fermi-LAT dwarf galaxies at present.

\section{Conclusions}

We entertained the possibility to communicate between Dirac fermion dark matter and the SM particles only through the four-form couplings to both the pseudo-scalar field and the Higgs field. The pseudo-scalar field reheats the Universe after the relaxation of the Higgs mass and it is responsible for making the dark matter in thermal equilibrium and undergoing the freeze-out process. The flux-induced mixing between the pseudo-scalar field and the Higgs field enables dark matter to annihilate into the SM particles without a velocity suppression while the direct detection bounds from XENON1T can be satisfied. 

There is a parameter space with a sizable Higgs mixing for explaining the relic density and accommodating the observable signals in Fermi-LAT and AMS-02, but subject to various existing bounds from Higgs-like scalar searches at the LEP, the LHC and Higgs and electroweak data from the LHC. In particular, it would be interesting to probe the bulk region of the parameter space where the relic density is determined dominantly by the dark matter annihilation into a pair of singlet-like scalars with similar mass as for dark matter, although being p-wave suppressed.
There are also resonance regions that are present in usual Higgs-portal dark matter models.

\section*{Acknowledgments}

The work is supported in part by Basic Science Research Program through the National Research Foundation of Korea (NRF) funded by the Ministry of Education, Science and Technology (NRF-2019R1A2C2003738 and NRF-2021R1A4A2001897). 
The work of YJK is supported in part by the National Research Foundation of Korea (NRF-2019-Global Ph.D. Fellowship Program). The work of AGM was supported by the Chung-Ang University Young Scientist Scholarship in 2019. The work of JS is supported by the Chung-Ang University Graduate Research Scholarship in 2020.

\def\theequation{A.\arabic{equation}}

\setcounter{equation}{0}

\vskip0.8cm
\noindent
{\Large \bf Appendix A: Scalar self-interactions }

From eq.~(\ref{sself}), we can also identify the scalar interactions for mass eigenstates as
\bea
{\cal L}_{\rm scalar, int} &=&  -\frac{1}{2} \mu c_H\, \varphi\, h^2 -\frac{1}{4}\lambda_{H,{\rm eff}} (v+h)^4  \nonumber \\
&=&-\frac{1}{2} \mu c_H\, (\cos\theta\,h_1 +\sin\theta\,h_2)   (-\sin\theta\,h_1 +\cos\theta\,h_2) ^2 \nonumber \\
&&- \frac{1}{4} \lambda_{H,{\rm eff}} (v-\sin\theta\,h_1 +\cos\theta\,h_2)^4 \nonumber \\
&=& -\kappa_{1} h^3_1-\kappa_2 h^3_2 -\kappa_{m1} h^2_1 h_2 -\kappa_{m2} h_1 h^2_2  \nonumber \\
&&-\frac{1}{4}\lambda_1 h^4_1-\frac{1}{4}\lambda_2 h^4_2 -\frac{1}{4} \lambda_{m1} h^3_1 h_2 -\frac{1}{4} \lambda_{m2}  h^2_1 h^2_2 -\frac{1}{4} \lambda_{m3} h_1 h^3_2
\eea
where
\bea
\kappa_1 &=& {1\over 2}\mu c_H \cos\theta \sin^2\theta- \lambda_{H,{\rm eff}}v \sin^3\theta,\\
\kappa_2 &=& {1\over 2}\mu c_H \sin\theta \cos^2\theta +\lambda_{H,{\rm eff}} v \cos^3\theta, \\
\kappa_{m1}&=& {1\over 2}\mu c_H \sin\theta(\sin^2\theta-2\cos^2\theta)+3\lambda_{H,{\rm eff}} v \cos\theta\sin^2\theta, \\
\kappa_{m2}&=& {1\over 2}\mu c_H \cos\theta (\cos^2\theta -2\sin^2\theta)-3\lambda_{H,{\rm eff}} v \sin\theta\cos^2\theta, \\
\lambda_1 &=& \sin^4\theta \lambda_{H,{\rm eff}}, \\
\lambda_2 &=& \cos^4\theta \lambda_{H,{\rm eff}}, \\
\lambda_{m1} &=& -4\cos\theta\sin^3\theta \lambda_{H,{\rm eff}},\\
\lambda_{m2} &=& 6\cos^2\theta\sin^2\theta \lambda_{H,{\rm eff}},\\
\lambda_{m3} &=& -4\sin\theta\cos^3\theta\lambda_{H,{\rm eff}}.
\eea

\def\theequation{B.\arabic{equation}}

\setcounter{equation}{0}

\vskip0.8cm
\noindent
{\Large \bf Appendix B: Formulas for scattering cross sections }

We list some of the exact formulas for annihilation and scattering cross sections for dark matter.

For the non-relativistic dark matter, the annihilation cross section for $\chi{\bar\chi}\to f{\bar f}$ is given by
\bea
(\sigma v_{\rm rel})_{\chi{\bar\chi}\to f{\bar f}} =\frac{1}{32\pi m^2_\chi}\, \Big(1- \frac{m^2_f}{m^2_\chi}\Big)^{1/2}\,\overline{|{\cal M}|^2} 
\eea
with
\bea
\overline{|{\cal M}|^2}_{\chi{\bar\chi}\to f{\bar f}}&=& \frac{1}{4} \left|\sum_{i=1,2}\,v_{f,i} \,\frac{{\bar v}_\chi(p_2)(v_{\chi,i}+ia_{\chi,i}\gamma^5)u_\chi(p_1)}{(p_1+p_2)^2-m^2_{h_i}} \right|^2 \bigg| {\bar u}_f(k_1) v_f(k_2)\bigg|^2 \nonumber \\
 &\simeq& 4 \Big[ 2(m^2_\chi-m^2_f) +\frac{1}{2} m^2_\chi v^2_{\rm rel} \Big]\bigg[\bigg(\sum_{i=1,2}\frac{v_{f_i} a_{\chi_i}}{4m^2_\chi-m^2_{h_i}}\bigg)^2\, \Big( 2m^2_\chi+\frac{1}{2} m^2_\chi v^2_{\rm rel}\Big) \nonumber \\
 &&+\frac{1}{2} m^2_\chi v^2_{\rm rel}\bigg(\sum_{i=1,2}\frac{v_{f_i} v_{\chi_i}}{4m^2_\chi-m^2_{h_i}}\bigg)^2\bigg] \nonumber \\
 &\simeq&\frac{m^2_f m^2_\chi}{v^2 f^2}\,(\sin 2\theta)^2 \Big[ 2(m^2_\chi-m^2_f) +\frac{1}{2} m^2_\chi v^2_{\rm rel} \Big]\bigg(\frac{1}{4m^2_\chi-m^2_{h_1}}-\frac{1}{4m^2_\chi-m^2_{h_2}}\bigg)^2 \nonumber \\
 &&\quad\times \bigg[ \Big( 2m^2_\chi+\frac{1}{2} m^2_\chi v^2_{\rm rel}\Big)\cos^2\beta+\frac{1}{2} m^2_\chi v^2_{\rm rel} \sin^2\beta \bigg].  \label{ff-full}
\eea
We used the above formula to get the approximate expression for a small velocity of dark matter in the text.

For the non-relativistic dark matter, the annihilation cross section for $\chi{\bar\chi}\to h_1 h_1 $ is also given by
\bea
( \sigma v_{\rm rel})_{\chi\bar{\chi}\rightarrow h_1 h_1}=( \sigma v_{\rm rel})^s+( \sigma v_{\rm rel})^{p1}+( \sigma v_{\rm rel})^{p2}+( \sigma v_{\rm rel})^{p3} \label{h1h1full}
\eea
with
\bea
&&( \sigma v_{\rm rel})^s={m_{\chi}^2 \cos^4\beta \sqrt{1-{m_{h_1}^2\over m_{\chi}^2}}\over 128 \pi f^4 (8m_{\chi}^4 -6m_{\chi}^2 m_{h_1}^2 + m_{h_1}^4)^2(4m_{\chi}^2-m_{h_2}^2)^2}  \nonumber \\
&&\times \bigg[ 4m_{\chi}^2(4m_{\chi}^2 - m_{h_1}^2)(4m_{\chi}^2-m_{h_2}^2)\cos^2\theta\sin 2\beta - f(2m_{\chi}^2-m_{h_1}^2)\sin^2\theta   \nonumber  \\
&&\quad\times \Big\{ c_H \mu (8m_{\chi}^2+m_{h_1}^2-3m_{h_2}^2)+3(m_{h_1}^2-m_{h_2}^2)\big( c_H \mu \cos 2\theta- 2\lambda_{H,{\rm eff}}v \sin 2\theta \big)  \Big\} \bigg]^2,
\eea
\bea
&&( \sigma v_{\rm rel})^{p1}=  {m_\chi^6 \cos^4\beta \cos^4\theta \sqrt{1-{m_{h_1}^2\over m_{\chi}^2}} \, v_{\rm rel}^2 \over 384\pi f^4(m_\chi^2-m_{h_1}^2)(2m_\chi^2-m_{h_1}^2)^4 }  \Big(24m_\chi^6-60m_\chi^4 m_{h_1}^2+54m_\chi^2 m_{h_1}^4-15m_{h_1}^6 \nonumber \\
&&\qquad\hspace{0.3cm} -8(8m_\chi^6-14m_\chi^4m_{h_1}^2+7m_\chi^2 m_{h_1}^4-m_{h_1}^6)\cos2\beta \nonumber \\
&&\qquad\hspace{0.3cm} +(56m_\chi^6-100m_\chi^4m_{h_1}^2+50m_\chi^2m_{h_1}^4-9m_\chi^6)\cos4\beta   \Big), 
\eea
\bea
&&( \sigma v_{\rm rel})^{p2}={m_\chi^4 \cos^3\beta\sin\beta \sin^2 2\theta  \sqrt{1-{m_{h_1}^2\over m_{\chi}^2}} \, v_{\rm rel}^2  \over 1536 \pi f^3 (m_\chi^2-m_{h_1}^2)(4m_\chi^2-m_{h_1}^2)^2(4m_\chi^2-m_{h_2}^2)^2(2m_\chi^2-m_{h_1}^2)^3} \nonumber \\
&&\times \bigg[ c_H\mu \Big( 3072m_\chi^{12}-256m_\chi^{10}(23m_{h_1}^2+9m_{h_2}^2)+32m_\chi^8(79m_{h_1}^4+154m_{h_1}^2m_{h_2}^2+9m_{h_2}^4)     \nonumber \\
&&\hspace{0.5cm} +16m_\chi^6m_{h_1}^2(15m_{h_1}^4-194m_{h_1}^2m_{h_2}^2-39m_{h_2}^4)\nonumber \\
&&\hspace{0.5cm}-4m_\chi^4m_{h_1}^4(76m_{h_1}^4-171m_{h_1}^2m_{h_2}^2-99m_{h_2}^4) \nonumber \\
&&\hspace{0.5cm}+2m_\chi^2m_{h_1}^6(14m_{h_1}^4-7m_{h_1}^2m_{h_2}^2-45m_{h_2}^4)-m_{h_1}^8m_{h_2}^2(m_{h_1}^2-3m_{h_2}^2) \Big) \nonumber \\
&&\hspace{0.3cm}+3(m_{h_1}^2-m_{h_2})\Big( 768m_\chi^{10}-32m_\chi^8(53m_{h_1}^2+3m_{h_2}^2)+16m_\chi^6m_{h_1}^2(75m_{h_1}^2+13m_{h_2}^2) \nonumber \\
&&\hspace{0.5cm}-12m_\chi^4m_{h_1}^4(28m_{h_1}^2+11m_{h_2}^2)+2m_\chi^2m_{h_1}^6(14m_{h_1}^2+15m_{h_2}^2)-m_{h_1}^8m_{h_2}^2  \Big) \nonumber \\
&&\hspace{0.5cm}\times (c_H\mu \cos2\theta-2\lambda_{\rm eff}v \sin2\theta) \nonumber \\
&&\hspace{0.3cm}+\cos2\beta \Big\{c_H\mu \Big( 8192m_\chi^{12}-128m_\chi^{10}(127m_{h_1}^2+43m_{h_2}^2)+128m_\chi^8(72m_{h_1}^4  \nonumber \\
&&\hspace{0.5cm}+91m_{h_1}^2m_{h_2}^2+6m_{h_2}^4) -8m_\chi^6m_{h_1}^2(148m_{h_1}^4+975m_{h_1}^2m_{h_2}^2+207m_{h_2}^4)\nonumber \\
&&\hspace{0.5cm} -32m_\chi^4 m_{h_1}^4(11m_{h_1}^4-61m_{h_1}^2m_{h_2}^2+36m_{h_2}^4) \nonumber \\ 
&&\hspace{0.5cm}+2m_\chi^2m_{h_1}^6(30m_{h_1}^4-49m_{h_1}^2m_{h_2}^2-159m_{h_2}^4)  -9m_{h_1}^8m_{h_2}^2(m_{h_1}^2-3m_{h_2}^2) \Big) \nonumber \\
&&\hspace{0.5cm} + 3(m_{h_1}^2- m_{h_2}^2)\Big( 1408m_\chi^{10}-256m_\chi^8(12m_{h_1}^2+m_{h_2}^2)+552m_\chi^6m_{h_1}^2(4m_{h_1}^2+m_{h_2}^2) \nonumber \\
&&\hspace{0.5cm} -128m_\chi^4m_{h_1}^6(30m_{h_1}^2+53m_{h_2}^2)-9m_{h_1}^8m_{h_2}^2  \Big)(c_H\mu \cos2\theta-2\lambda_{\rm eff}v \sin2\theta) \Big\}  \bigg], 
\eea
\bea
&&( \sigma v_{\rm rel})^{p3}={m_\chi^2\cos^2\beta \sin^4\theta \sqrt{1-{m_{h_1}^2\over m_{\chi}^2}} \, v_{\rm rel}^2 \over 1024  \pi f^2(m_\chi^2-m_{h_1}^2)(4m_\chi^2-m_{h_1}^2)^3(4m_\chi^2-m_{h_2}^2)^3} \Big( c_H\mu(8m_\chi^2+m_{h_1}^2-3m_{h_2}^2)\nonumber \\
&&\quad\hspace{0.3cm} +3(m_{h_1}^2-m_{h_2}^2)(c_H\mu \cos2\theta-2\lambda_{\rm eff}v \sin2\theta) \Big)\nonumber \\
&&\quad\hspace{0.3cm}\times \Big\{32c_H\mu m_\chi^2(4m_\chi^2-m_{h_2}^2)(4m_\chi^4-5m_\chi^2m_{h_1}^2+m_{h_1}^4)\cos^2\beta \nonumber \\
&&\quad\hspace{0.5cm}+ \Big( 2(4m_\chi^2-m_{h_2}^2)(4m_\chi^4-5m_\chi^2 m_{h_1}^2+m_{h_1}^4)-\big( 160m_\chi^6-8m_\chi^4(25m_{h_1}^2+3m_{h_2}^2) \nonumber \\
&&\quad\hspace{0.5cm}+2m_\chi^2 m_{h_1}^2(14m_{h_1}^2+15m_{h_2}^2)-3m_{h_1}^4m_{h_2}^2  \big)\cos^2\beta \Big) \nonumber\\
&&\quad\hspace{0.3cm}\times\Big(c_H\mu (8m_\chi^2+m_{h_1}^2-3m_{h_2}^2)+3(m_{h_1}^2-m_{h_2}^2)( c_H \mu \cos 2\theta- 2\lambda_{\rm eff}v \sin 2\theta) \Big)  \Big\}.
\eea
We used the above formula to get the approximate expression in the limit of a small Higgs mixing angle in the text.

For the direct detection of dark matter, we also derive the cross section for the spin-independent elastic scattering between dark matter and nucleus, as follows,
 \bea
\sigma_{\chi-N} &=&\frac{1}{16\pi (m_\chi+m_N)^2}\, \overline{|{\cal M}|^2}_{\chi N\to \chi N}
 \eea
 with
\bea
\overline{|{\cal M}|^2}_{\chi N\to \chi N}  =\frac{4m^2_\chi m^2_N}{v^2A^2} \bigg[ \bigg( \sum_{i=1,2}  \frac{ {\tilde v}_i v_{\chi,i}}{m^2_{h_i}}\bigg)^2+ \frac{{\vec q}^2}{4m^2_\chi} \,\bigg(\sum_{i=1,2}\frac{{\tilde v}_i a_{\chi,i}}{m^2_{h_i}}\bigg)^2\bigg] \Big(Z f_p + (A-Z) f_n\Big)^2\bigg|_{{\vec q}^2=2m_N E_R}.
\eea
Here,  we note that the momentum transfer is taken to give the recoil energy $E_R$ for the nucleus after the scattering. Thus, the momentum dependent term is suppressed by $m_N E_R/m^2_\chi$, which is less than $6\times 10^{-7}$ for $m_N\simeq 131 m_p$ for XENON1T, $E_R\lesssim 50\,{\rm keV}$ and $m_\chi\sim 100\,{\rm GeV}$.
Therefore, for $\sum_{i=1,2}\frac{{\tilde v}_i v_{\chi,i}}{m^2_{h_i}}=0$, the elastic scattering cross section between dark matter and nucleus becomes suppressed by the momentum transfer.

\end{document}